\documentclass[letterpaper,10pt]{article}
\usepackage[round]{natbib}

\usepackage[margin=.75in]{geometry}
\usepackage[T1]{fontenc}
\usepackage{fix-cm}
\usepackage[utf8]{inputenc}
\usepackage{amsmath, amssymb, amsfonts}
\usepackage{mathtools}
\usepackage{amsthm}
\usepackage{bm}
\usepackage[hidelinks]{hyperref}
\usepackage{algorithm}
\usepackage{algorithmic}
\usepackage{xfrac}
\usepackage[labelformat=simple]{subcaption}

\graphicspath{{figures/}}

\hypersetup{pdftitle={Generic second-order macroscopic traffic node model for general multi-input multi-output road junctions via a dynamic system approach}}

\theoremstyle{definition}
\newtheorem{defn}{Definition}

\theoremstyle{remark}
\newtheorem{rem}{Remark}

\theoremstyle{plain}

\newcommand{\etab}{\bm{\eta}}

\newcommand{\RR}{\mathbb{R}}

\title{Generic second-order macroscopic traffic node model for general multi-input multi-output road junctions via a dynamic system approach}
\author{Matthew A. Wright and Roberto Horowitz}
\date{}

\begin{document}

\maketitle

\begin{abstract}
This paper addresses an open problem in traffic modeling: the second-order macroscopic node problem.
A second-order macroscopic traffic model, in contrast to a first-order model, allows for variation of driving behavior across subpopulations of vehicles in the flow.
The second-order models are thus more descriptive (e.g., they have been used to model variable mixtures of behaviorally-different traffic, like car/truck traffic, autonomous/human-driven traffic, etc.), but are much more complex.
The second-order node problem is a particularly complex problem, as it requires the resolution of discontinuities in traffic density and mixture characteristics, and solving of throughflows for arbitrary numbers of input and output roads to a node (in other words, this is an arbitrary-dimensional Riemann problem with two conserved quantities).
In this paper, we extend the well-known ``Generic Class of Node Model'' constraints to the second order and present a simple solution algorithm to the second-order node problem.
Our solution makes use of a recently-introduced dynamic system characterization of the first-order node model problem, which gives insight and intuition as to the continuous-time dynamics implicit in node models.
We further argue that the common ``supply and demand'' construction of node models that decouples them from link models is not suitable to the second-order node problem.
Our second-order node model and solution method have immediate applications in allowing modeling of behaviorally-complex traffic flows of contemporary interest (like partially-autonomous-vehicle flows) in arbitrary road networks.
\end{abstract}

\section{Introduction}
\label{gsom_node:sec:intro}
Road congestion is a well-known driver of inefficiency and waste in contemporary societies.
According to one recent study, congestion caused drivers to waste more than 3 billion gallons of fuel and nearly 7 billion extra hours in 2015 in the U.S. alone \citep{schrank_2015_2015}.
Building an understanding of the emergent, macro-scale dynamics of traffic flow that result from many vehicles interacting is an important step towards intervening to lessen these costs.
This is especially true today, as emerging technologies such as connected and autonomous vehicles and smart infrastructure allow transportation engineers and operations researchers finer and more pervasive control to mitigate congestion \citep{ge_dynamics_2014,cui_stabilizing_2017,wu_emergent_2017,stern_dissipation_2018}.

The macroscopic approximation of vehicle traffic has proven a valuable tool for traffic modeling and control applications. 
This macroscopic theory describes the dynamics of vehicles along roads with partial differential equations (PDEs) inspired by fluid flow.
The most basic macroscopic formulation is the so-called ``kinematic wave'' or ``Lighthill-Whitham-Richards'' (LWR) model due to \citet{lighthill_kinematic_1955_pt1,lighthill_kinematic_1955_pt2} and \citet{richards_shock_1956}, which describes traffic with a one-dimensional conservation equation,
\begin{equation}
  \frac{\partial \rho}{ \partial t} + \frac{ \partial (\rho v)}{\partial x} = 0 \label{prelim:eq:LWR}
\end{equation}
where $\rho(x,t)$ is the density of vehicles, $t$ is time, $x$ is the lineal direction along the road, and $v(\rho)$ is the flow speed.
The total flow, $\rho v$, is often expressed in terms of a flux function, $f(\rho) = \rho v$ (the flux function on a long straight road is often called the \emph{fundamental diagram}).

For many traffic studies, the LWR-type formulation \eqref{prelim:eq:LWR} can be used to model a number of traffic dynamics of interest.
For example, in dynamic traffic assignment studies, modeled vehicle demands are fed into a road network model, and a first-order dynamical simulation of the resulting network flows can estimate -- among other items of interest that result from the nonlinear dynamics -- where traffic jams will appear and how long they will last, and quantify the general efficiency of the road network as a function of the input demands and the network geometry and topology.

However, there are some macroscopic traffic phenomena that an LWR-type formulation \eqref{prelim:eq:LWR} cannot capture.
Three particular examples that have been discussed in the literature are hysteresis loops in the $(\rho, v)$ plane (e.g., \citet{treiterer_hysteresis_1974,zhang_mathematical_1999}), drops in capacity when roads become congested \citep{hall_capacity_1990}, and the occasional emergence of congestion behavior in regions of low density and no bottlenecks (that is, in conditions where we should expect free-flow behavior) \citep{sugiyama_traffic_2008}.
These phenomena cannot be captured in solutions to the LWR-type model, so, in situations where these smaller-scale (that is, relative to ``larger-scale'' phenomena like bottlenecks) macroscopic phenomena are of importance to the modeler, more expressive models that can produce them are appropriate. See, e.g., \citet{wong_multi-class_2002,zhang_non-equilibrium_2002} for some examples where the LWR-type models' lack of expressiveness are used as justifications for extensions.

One extension of the LWR model that can express a richer variety of dynamics is the so-called Aw-Rascle-Zhang (ARZ) \citep{aw_resurrection_2000, zhang_non-equilibrium_2002} family of models.
These models fit into the so-called ``generic second order''\footnote{
As seen in \eqref{eq:gsom} and previously pointed out by many authors, the ``second-order'' model actually consists of two first-order partial differential equations (that is, they only contain first derivatives).
In a case of overloaded mathematical terminology, the name ``second order'' here comes from a system-theoretic view, where a second-order system is one that has two state variables: in this case, $\rho$ and $w$ (or, equivalently, $\rho$ and $v$).}
or ``extended ARZ'' class of traffic models \citep{lebacque_generic_2007}, which can be written as
\begin{subequations}
\begin{align}
  \frac{\partial \rho}{ \partial t} + \frac{ \partial (\rho v)}{\partial x} &= 0 \label{eq:gsom1} \\
  \frac{\partial w}{ \partial t} + v \frac{ \partial w}{\partial x} &= 0 \label{eq:gsom2} \\
  \textnormal{where } v &= V(\rho, w) \label{eq:gsom3}
\end{align}
\label{eq:gsom}
\end{subequations}
where $w(x,t)$ is a property or invariant that is conserved along trajectories \citep{lebacque_generic_2007}.
In words, \eqref{eq:gsom1} is a continuity equation of $\rho$, \eqref{eq:gsom2} is an advection equation of $w$, and \eqref{eq:gsom3} defines the velocity field that governs both the flux of $\rho$ and the speed at which $w$ moves through space.

The property $w$ in \eqref{eq:gsom} can be described as a characteristic of vehicles that determines their density-velocity relationship.
Members of the generic second order model (GSOM) family are differentiated by the choice of $w$ and its relationship on the $\rho$-$v$ behavior.
Examples of properties modeled by chosen $w$'s include the difference between vehicles' speed and an equilibrium speed \citep{aw_resurrection_2000,zhang_non-equilibrium_2002}, driver's desired spacing \citep{zhang_non-equilibrium_2002}, or the flow's portion of autonomous vehicles \citep{wang_comparing_2017}.
An intuitive way of describing the effect of the property $w$ in \eqref{eq:gsom} is that it parameterizes a family of flow models, $f(\rho, w) = \rho V(\rho, w)$, with different flow models for different values of $w$ \citep{lebacque_generic_2007, fan_collapsed_2017}.
In particular, different classes of vehicles (e.g., autonomous vs. human-driven) are assumed to have different equations governing their driving behavior: these varying dynamics are aggregated to create an averaged dynamics in the macroscopic flow equation \eqref{eq:gsom1}.
The aggregate dynamical behavior (parameterized by $w$) of course then tracks the vehicles themselves, as reflected in \eqref{eq:gsom}.

The models covered thus far describe traffic along single roads.
For application to multiple roads with intersections, these road networks are often modeled as directed graphs.
Edges that represent individual roads are called links, and junctions where links meet are called nodes.
Typically, the flow model $f(\cdot)$ on links is called the ``link model,'' and the flow model at nodes is called the ``node model.''
Development of accurate link and node models have been areas of much research activity in transportation engineering for many years.

This paper focuses on node models for first- and second-order macroscopic models.
The node model resolves the discontinuities in $\rho$ and/or $w$ between links and determines a flux boundary condition at each link's exit (for incoming links) or entrance (for outgoing links).
In other words, the node model takes as input the Dirichlet boundary condition for each link at the junction, and outputs a resulting Neumann boundary condition.
For nodes with merges, diverges, or both, this Riemann problem becomes multidimensional.
Through this, the node model determines how the state of an individual link affects and is affected by its connected links, their own connected links, and so on through the network.
As a result, it has recently been recognized that the specific node model used can have a very large role in describing the network-scale congestion dynamics that emerge in complex and large networks (for more on this, see the discussions in, e.g., the introduction sections of \citet{tampere_generic_2011} and \citet{jabari_node_2016}).

In \citet{wright_dynamic_2016}, we introduced a novel characterization of node models as dynamic systems.
Traditional studies of node models (see, e.g., \citet{tampere_generic_2011, flotterod_operational_2011, corthout_non-unique_2012, smits_family_2015, jabari_node_2016, wright_node_2017}) usually present the node model as an optimization problem (where the node flows are found by solving this problem) or in algorithmic form (where an explicit set of steps are performed to compute the flows across the node).
In contrast, the dynamic system characterization describes the flows across the node as themselves evolving over some period of time (in application, this means that the dynamic system characterization presents time-varying dynamics that are said to occur during the simulation timesteps of the link PDEs).
The dynamic system characterization can be thought of as making explicit the time-varying behavior of the flows at nodes of many algorithmic node models: it was shown in \citet{wright_dynamic_2016} that the dynamic system characterization produces the same solutions as the algorithm introduced in \citet{wright_node_2017}, which also reduces to the one introduced in \citet{tampere_generic_2011} as a special case.

The dynamic system characterization has proven useful in imparting an intuition as to what physical processes over time are implicit in these algorithmic node models (see the discussions referring to \citet{wright_dynamic_2016} in \citet{wright_node_2017} for some examples).
In this paper, we develop a dynamic system characterization of a second-order node model, and use it to solve the general node problem for second-order models.

This paper has several main contributions.
The first is an extension of the dynamic system characterization of first-order node models as introduced in \citet{wright_dynamic_2016} to a simple, closed-form solution algorithm.
This represents the completion of an argument began in Section 4.1 of that reference.
The second contribution is the extension of the dynamic system characterization to the generic second-order models.
As we will see, the dynamic system characterization lends itself to an intuitive incorporation of the second PDE in \eqref{eq:gsom} that is not obvious in the traditional, optimization-problem presentation of node models.
The third contribution, and the principal contribution of this paper, parallels the first by using the second-order dynamic system node model to derive an intuitive, closed-form algorithm for computing node flows for second-order flow models for general, multi-input multi-output nodes.
To the best of our knowledge, this represents the first proposed generic (applicable to multi-input multi-output nodes with arbitrary numbers of input and output links) node flow solver for second-order traffic flow modeling.\footnote{
A note on naming: as we will see in Section \ref{sec:gcnmfirstorder}, we build off the so-called ``generic class of first-order node models'' to develop our second-order node model. Given that the relevant second-order model used \eqref{eq:gsom} is itself called the ``generic second order model,'' it might be accurate to describe this paper's results as the ``genericization of the generic class of node models to the generic second-order model,'' but this description likely loses in comprehensibility what it might gain in accuracy.}
Finally, our fourth contribution is an argument that the second-order node problem is not well-suited to the pervasive ``supply and demand'' CTM-like discretization that is highly prevalent in macroscopic modeling (see Section \ref{sec:secondorder_discussion}).

The remainder of this paper is organized as follows.
Section \ref{sec:firstorder} reviews the first-order node flow problem, the first-order dynamic system characterization introduced in \citet{wright_dynamic_2016}, and presents the aforementioned closed-form solution algorithm (contribution one in the above paragraph).
Section \ref{sec:secondorder} reviews the link discretization of the GSOM \eqref{eq:gsom} as presented in, e.g., \citet{lebacque_generic_2007, fan_collapsed_2017}, which produces the inputs to our second-order node model, and the standard one-input one-output second-order flow problem and its solution.
Section \ref{sec:secondordernode} presents the extension of the second-order flow problem to the multi-input multi-output case, the dynamic system characterization to the GSOM family \eqref{eq:gsom} and the solution algorithm for the general node problem (contributions two and three).
Finally, Section \ref{sec:conclusion} concludes and notes some open problems.

\section{First-order node model}
\label{sec:firstorder}
In this section, we review the general first-order node problem and a particular node model (and its solution algorithm).
This node model will be extended to the second-order node problem in section \ref{sec:secondorder}.

The traffic node problem is defined on a junction of $M$ input links, indexed by $i$, and $N$ output links, indexed by $j$.
We further define $C$ ``commodities'' of vehicle, indexed by $c$.
Each commodity $c$ denotes a different kind of vehicle (e.g., cars, trucks, autonomous vehicles, etc.).
\footnote{Commodities are sometimes called vehicle classes. In this work, we purposefully use the term ``commodity'' when referring to different types of vehicles to differentiate from the usage of the term ``class'' with respect to node models.}
The first-order node problem takes as inputs the incoming links' per-commodity demands $S_i^c$, split ratios $\beta_{i,j}^c$ (which define the portion of vehicles of commodity $c$ in link $i$ that wish to exit to link $j$), and outgoing links' supplies $R_j$, and gives as outputs the set of flows from $i$ to $j$ for commodity $c$, $f_{i,j}^c$.
We denote as a shorthand the per-commodity directed demand $S_{i,j}^c \triangleq \beta_{i,j}^c S_i^c$.
Nodes are generally infinitesimally small and have no storage, so all the flow that enters the node must exit the node.

The rest of this section is organized as follows.
Section \ref{sec:gcnmfirstorder} defines our first-order node problem as an optimization problem defined by explicit requirements, following the example set by \citet{tampere_generic_2011}.
Section \ref{sec:firstorderds} reviews the dynamic system of \citet{wright_dynamic_2016} whose executions produce solutions to the node problem.
Finally, section \ref{sec:firstorder_algorithm} uses the dynamic system formulation as a base to develop a node model solution algorithm.
This algorithm represents the completion of an argument began in \citet{wright_dynamic_2016}.

\subsection{``Generic Class of Node Model'' requirements}
\label{sec:gcnmfirstorder}
The node problem's history begins with the original formulation of macroscopic discretized first-order traffic flow models \citet{daganzo_cell_1995}.
There have been many developments in the node model theory since, but we reflect only some more recent results.

We can divide the node model literature into pre- and post-\citet{tampere_generic_2011} epochs.
They drew from the literature several earlier-proposed node model requirements to develop a set of conditions for first-order nodel models they call the ``generic class of first-order node models'' (GCNM).
These set of conditions give an excellent starting point for our discussion of the mathematical technicalities of node models, and have been used as a starting point by many subsequent papers, such as \citet{flotterod_operational_2011, corthout_non-unique_2012, smits_family_2015, jabari_node_2016, wright_node_2017}.
In the following list, we present the variant of first-order GCNM requirements used in \citet{wright_node_2017}, which includes a modification of the first-in-first-out (FIFO) requirement (item \ref{item:fifo} below) to \citet{wright_node_2017}'s ``partial FIFO'' requirement.

\begin{enumerate}
	\item Applicability to general numbers of input links $M$ and
	output links $N$. In the case of multi-commodity flow, this
	also extends to general numbers of commodities $c$.
	
	\item Maximization of the total flow through the node.
	Mathematically, this may be expressed as $\max \sum_{i,j,c} f_{i,j}^c$.
	According to~\citet{tampere_generic_2011}, this means that
	``each flow should be actively restricted by one of the constraints,
	otherwise it would increase until it hits some constraint.''
	When a node model is formulated as a constrained optimization problem,
	its solution will automatically satisfy this requirement.
	However, what this requirement really means is that constraints should be
	stated \emph{correctly} and not be overly simplified (and thus,
	overly restrictive) for the sake of convenient problem formulation.
	See the literature review in \citet{tampere_generic_2011} for examples of node models
	that inadvertently do not maximize node throughput by oversimplifying their
	requirements. \label{item:maxFlow}
	
	\item Non-negativity of all input-output flows. Mathematically, $f_{i,j}^c \ge 0$
	for all $i,j,c$.
	
	\item Flow conservation: Total flow entering the node must be equal to total flow
	exiting the node. Mathematically, $\sum_i f_{i,j}^c = \sum_j f_{i,j}^c$ for all $c$.
	\item Satisfaction of demand and supply constraints. Mathematically, $\sum_j f_{i,j}^c \le S_i^c$ and $\sum_i f_{i,j}^c \le R_j$.
	
	\item Satisfaction of the (partial) first-in-first-out (FIFO) constraint:	if a single
	destination $j'$ for a given $i$ is not able to accept all demand
	from $i$ to $j'$, then all other flows from $i$ are constrained by the queue of
	$j'$-destined vehicles that builds up in $i$. \label{item:fifo}

	We will spend some time outlining the specifics of the partial FIFO constraint and developing an intuition for it in section \ref{sec:partialfifo}.
	
	\item Satisfaction of the invariance principle.
	This principle was introduced by \citet{lebacque_first-order_2005} and specifies that ``under constant demand and supply constraints, flows should be invariant during an infinitesimal timestep.'' \citep{tampere_generic_2011}.
	In particular, this requirement means that the portioning of a supply $R_j$ among the input links $i$ not be proportional to the demands $S_i$.
	
	\item Supply restrictions on a flow from any given input link are
imposed on commodity components of this flow proportionally to their 
per-commodity demands. Mathematically, $f_{i,j}^c/(\sum_c f_{i,j}^c) = \beta_{i,j}^c S_{i}^c/(\sum_c \beta_{i,j}^c S_{i}^c)$. 

This assumes that the commodities are mixed isotropically.
This means that all vehicles attempting to take movement $i,j$ will be queued in roughly random order, and not, for example, having all vehicles of commodity $c=1$ queued in front of all vehicles of $c=2$, in which case the $c=2$ vehicles would be disproportionally affected by spillback.
We feel this is a reasonable assumption for situations where the demand at the node is dependent mainly on the vehicles near the end of the link (e.g., in a small cell at the end).\label{item:classprop}

\item A ``supply constraint interaction rule'' (SCIR), the idea of which was explicitly defined by \citet{tampere_generic_2011}.
We discuss this in more detail in section \ref{sec:scir}.
\end{enumerate}

\subsubsection{Details of the Partial FIFO constraint}
\label{sec:partialfifo}
In this work, we use the ``partial'' FIFO constraint, item \ref{item:fifo} above, which was introduced in \citet{wright_node_2017}.
A classic (non-partial) FIFO constraint (also referred to as a ``conservation of turning fractions'' constraint by \citet{tampere_generic_2011}) has the mathematical form $\beta_{i,j}^c = f_{i,j}^c / (\sum_j f_{i,j}^c)$, and means that when some vehicles that want to go to a particular $j$ are unable to take this movement, they queue and block all vehicles behind them, regardless of whether the blocked vehicles want to go to that particular $j$ or another one.

In \citet{wright_node_2017}, we argued that this full FIFO requirement can be unrealistic if like $i$ has multiple lanes, since it necessarily implies that all lanes are blocked, by vehicles queueing for output $j$ even if the movement $(i,j)$ is only accessible by a subset of the lanes (supposing of course that vehicles queueing for $j$ will not be queueing on lanes that they cannot use to reach $j$).
The partial FIFO requirement encodes how vehicles queueing for a movement $(i,j')$ will only partially block another movement $(i,j)$, with the degree of blockage being related to the degree that $(i,j)$'s lanes are blocked by this queue.
In our model, we will say that one of these partial blockages affects the blocked flow by reducing the movement's capacity $F_{i,j}$, where the nominal (un-reduced) capacity is defined as the maximum possible flow for the particular movement.\footnote{The constraint $\sum_c f_{i,j}^c \leq F_{i,j}$, that the flow be no more than the maximum possible flow (before any capacity reductions due to blockages) is generally not given in node model problem statements because it is implied by demand constraint, since by definition $\sum_c \beta_{i,j}^c S_{i}^c \leq F_{i,j}$.}
Following, e.g., \citep{tampere_generic_2011}, we will define the movement capacity $F_{i,j}$ as a portion of the input link capacity $F_i$ weighted by the vehicles that are making use of that capacity,
\begin{equation}
	F_{i,j} = \frac{\sum_c S_{i,j}^c}{\sum_c S_i^c} F_i. \label{eq:directed_capacity}
\end{equation}

To move to specifics, suppose a particular link $j'$ is not able to receive all the vehicles from $i$ that want to enter it, and the denied vehicles queue up.
The degree to which this queue restricts the other flows $f_{i,j}^c$ is partially 
defined by
\emph{restriction intervals} $\etab_{j',j}^i = [y,z] \subseteq [0,1]$.
This interval means that a queue in the $i,j'$ movement will block the portion of
$i,j$-serving lanes in $i$ with leftmost extent $y$ and rightmost extent $z$ (e.g.,
if $i,j$ is a through movement that uses two lanes and $i,j'$ is a right-turn
movement that uses the right of those two lanes, then 
$\etab_{j',j}^i = [\sfrac{1}{2}, 1]$.\footnote{
Continuing this example, we will have $\etab_{j,j'}^i = [0,1]$ since the only lane
in $i$ that serves movement $i,j'$ (the right lane) 
will be blocked by a queue for the through movement, which will queue on both lanes).}
The traditional, full FIFO behavior, where any queue in $i$ blocks all of $i$'s
lanes, can be recovered by setting all $\etab_{j',j}^i = [0,1]$.
	\footnote{To help keep the meaning of $\etab_{j',j}^i$ clear, we find it helpful to read it as ``the restriction interval of $j'$ onto $j$ for $i$''}
	
\begin{figure}[htb]
	\centering
	\begin{subfigure}[b]{.45\linewidth}
		\centering
		\includegraphics[width=\linewidth]{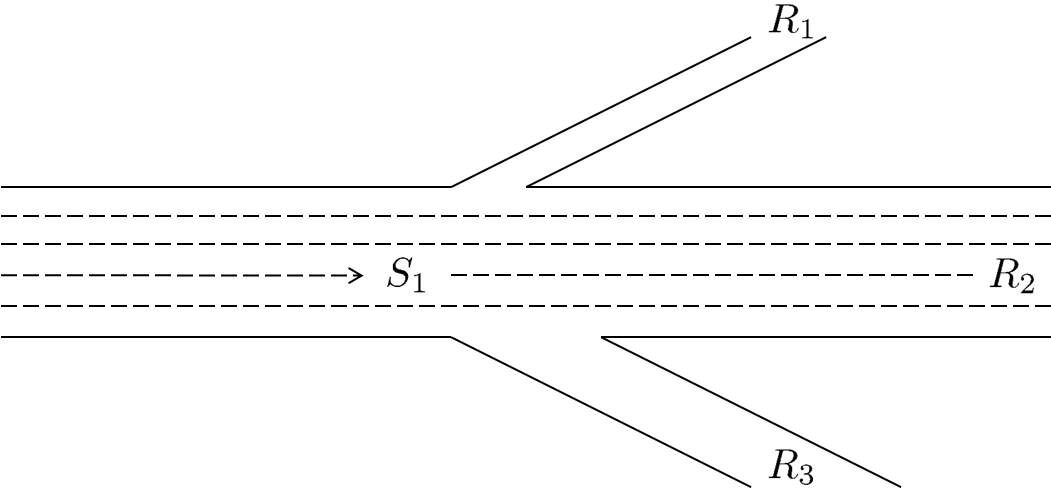}
		\caption{}
		\label{fig:fifo_example_road}
	\end{subfigure}
	\begin{subfigure}[b]{.45\linewidth}
		\centering
		\includegraphics[width=\linewidth]{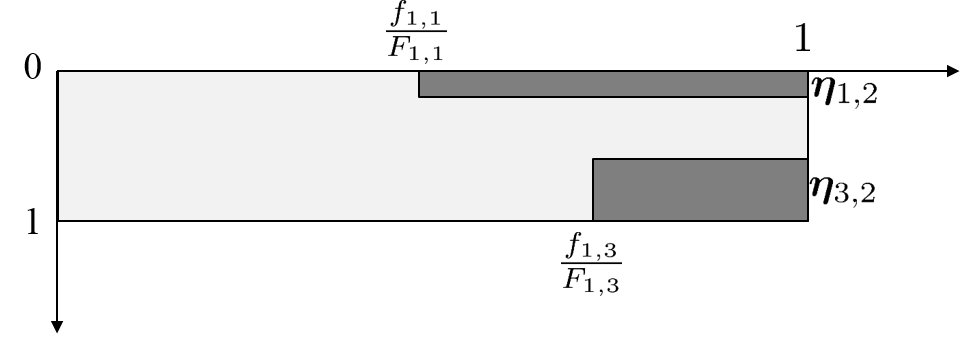}
		\caption{}
		\label{fig:fifo_example_rectangle}
	\end{subfigure}
	\caption[Example application of partial FIFO.]{Example application of partial FIFO. \subref{fig:fifo_example_road}: A one-input, three-output road junction. \subref{fig:fifo_example_rectangle}: Pictorial representation of restriction intervals on $f_{1,2}$: this flow is upper-bounded by the nominal capacity of movement $(1,2)$ multiplied by the lightly-shaded portion of the rectangle.
	The dark-shaded portions represent the capacity that is blocked by queueing vehicles in the partial FIFO construction.}
	\label{fig:fifo_example}
\end{figure}

Figure \ref{fig:fifo_example} illustrates the application of partial FIFO in a slightly more complicated example.
This example was previously discussed in \citet{wright_dynamic_2016,wright_node_2017}.
In Figure \ref{fig:fifo_example_road}, a five-lane road diverges into three output links.
For the sake of this example, we say that only the leftmost lane can access movement $(1,1)$, all five lanes can access movement $(1,2)$, and only the two rightmost lanes can access movement $(1,3)$.
So, the restriction intervals for this example, for the restrictions onto $(1,2)$'s flow, are $\etab_{1,2}^1 = [0, \sfrac{1}{5}]$, $\etab_{2,2}^1 = [0, 1]$, and $\etab_{2,3}^1 = [\sfrac{3}{5}, 1]$.

Figure \ref{fig:fifo_example_rectangle} shows a graphical computation of the effects of restriction intervals on movement $(1,2)$'s capacity, given particular flows for the other movements $f_{1,1}$ and $f_{1,3}$.
In this particular example, we are supposing that both $f_{1,1}$ and $f_{1,3}$ are supply-constrained (i.e., $f_{1,1}=R_1$ and $f_{1,3}=R_3$).
The dark-shaded portions of the figure show the effects of the vehicles queueing for these two movements on the $(1,2)$ movement's capacity.
The extent of the dark-shaded regions on the vertical axis are exactly the restriction intervals $\etab_{j',2}^1$ for $j' \in \{1,3\}$.

We now explain the meaning of the horizontal-axis extent of the dark-shaded regions.
By definition, the capacity is the maximum possible amount of vehicles that can pass through a movement in some time period.
So, one minus the ratio of a supply-constrained movement flow to the movement's capacity $(1 - \sfrac{f_{i,j}}{F_{i,j}})$ is then the portion of the time period for which $F_{i,j}$ is defined\footnote{Recall the units of capacity $F$ are $\sfrac{\textnormal{veh}}{\textnormal{time}}$} that the movement $(i,j)$ is blocked by a queue.
So, returning to Figure \ref{fig:fifo_example_rectangle}'s example, the restriction intervals reduce the capacity of the $(1,2)$, proportional both to the intervals' length and the of the ``portion of time'' that their queue exists.

Putting together the pieces, the (partial) FIFO requirement can be stated mathematically as
\begin{equation}
f_{i,j}^c \leq F_{i,j} \frac{S_{i,j}^c}{\sum_c S_{i,j}^c} \left(1 - \mathcal{A} \left( \bigcup_{j' \neq j} 
	\left\{ \bm{\eta}_{j',j}^i \! \times \! \left[ \frac{f_{i,j'}}{F_{i,j'}} , \, 1 \right] \right\} \right)
	\right) \label{eq:rectangleRelaxedFifo}
\end{equation}
where $\mathcal{A}(\cdot)$ denotes the area of a two-dimensional object, $\times$ denotes a Cartesian product, $F_{i,j}$ is given by \eqref{eq:directed_capacity}, and $f_{i,j} \triangleq \sum_c f_{i,j}^c$.
Note that we take the union of the two-dimensional objects that define the extent of the partial FIFO restriction (the dark-shaded regions in Figure \ref{fig:fifo_example_rectangle}) to obtain the effective partial FIFO restriction of \emph{all} blocking queues.

The formulation in \eqref{eq:rectangleRelaxedFifo} is complicated in order to state it
as an optimization constraint.
The time-period intuition developed above will be used explicitly in the dynamic system definition of the node models.
A major contribution of the dynamic system approach to node modeling is the
explicit encoding of this more intuitive description.

\paragraph{Note:} In prior work on the partial FIFO constraint, \citet{wright_node_2017,wright_dynamic_2016}, our construction was slightly different than as stated in \eqref{eq:rectangleRelaxedFifo} above.
In particular, the quantity that was brought down as restriction intervals activate in the prior version was the \emph{demand} rather than the capacity.
As we discuss in \citet{wright_macro-scale_2019}, this prior version leads to unintentional unrealistic constraints on the flow that are rectified by tightening the capacity constraint instead of the demand constraint.

\subsubsection{The ``Supply Constraint Interaction Rule''}
\label{sec:scir}
The term ``supply constraint interaction rule'' (SCIR), introduced by \citet{tampere_generic_2011}, refers to two other elements needed
to define a node model.
The first element is a rule for the portioning of output link supplies
$R_j$ among the input links. Following~\citet{gentile_spillback_2007}, in~\citet{tampere_generic_2011} it was 
proposed to allocate supply for incoming flows proportionally to input link capacities $F_i$.
In this work, we allocate supply proportionally to the links' ``priorities'' $p_i$ (in the spirit of \citet{daganzo_cell_1995,ni_simplified_2005,flotterod_operational_2011, wright_node_2017}).
In the dynamic system view, priorities represent the relative rate at which vehicles exit each link $i$ to claim downstream space (one reasonable formulation might be to follow the capacity-proportional example, $p_i = F_i$, if, as in \citet{tampere_generic_2011}, it assumed that vehicles exit a link at rate $F_i$).
In this work, we assume for simplicity that all $p_i > 0$.\footnote{
	In a more general formulation, a priority of $p_i = 0$ makes sense in a ``staged'' scheme, where a zero-priority link or movement only gets to send its vehicles after other input links have exhausted their demand.
	As mentioned in \cite[Section 4.2]{wright_node_2017}, this is equivalent to a common assumption in freeway onramp junction modeling (particularly models for control schemes like ALINEA \citep{papageorgiou_alinea_1991}) the onramp gets to fill all its demand, and the freeway gets the remainder.
	This staging of priorities is also related to \citet{jabari_node_2016}'s suggestion of modeling signalized junctions such that protected movements get to claim all the supply they can use before non-protected movements.
	One can fit these staged-priority problems into the node problem scheme presented here by breaking them into sub-problems that are solved sequentially, with different sets of links having nonzero priority for each sub-problem.
}

The second necessary element is a redistribution of ``leftover supply.'' Following the
initial partitioning of supplies $R_j$, if one or more of the supply-receiving input links
does not fill its allocated supply,
some rule must redistribute the difference to other input links who may still fill it.
This second element is meant to model the selfish behavior of drivers to take any space
available, and ties in closely with requirement \ref{item:maxFlow} above.
\citet{tampere_generic_2011} referred to these two elements collectively as a ``supply
constraint interaction rule'' (SCIR).
For some discussion of choices of SCIRs in recent papers, see \citet[Section 2.1]{wright_node_2017}, and for more background discussion of the need for an SCIR, see \citet[Sections 2 and 4]{tampere_generic_2011}.

In this work, we consider a SCIR of the form (equation (3.37) in \citet{wright_node_2017}, but rearranged slightly)
\begin{subequations} \label{eq:scir}
\begin{align}
		& \sum_j f_{i,j} < \sum_c S_i^c \implies W_i \neq \emptyset \quad \forall \, i \label{eq:scir1} \displaybreak[0] \\
  		& f_{i,j} \geq \frac{
	p_{i,j}}{  \sum_{i' = 1}^M p_{i',j}} R_j \quad \forall \, j \in W_i \quad
	\forall i \label{eq:scir2} \displaybreak[0]
\end{align}
where
\begin{equation}
W_i = \left\{ j^\ast : \;
\sum_c \beta_{i,j^\ast}^c S_i^c > 0 \text{ and } \nexists \; i' \neq i \text{ s.t. } \frac{f_{i,j^\ast}}{p_{i,j^\ast}} < \frac{f_{i',j^\ast}}{p_{i',j^\ast}}\right\} \label{eq:scirw}
\end{equation}
\end{subequations}
and
\begin{equation}
  p_{i,j} = \frac{\sum_c S_{i,j}^c}{\sum_c S_i^c} p_i \label{eq:orientedpriority}
\end{equation}
is the ``oriented priority,'' which distributes input links' priority proportionally to the actual vehicles using that priority to claim downstream supply, and $f_{i,j} \triangleq \sum_c f_{i,j}^c$ (as before).
In the case of capacity-equivalent priorities $p_i=F_i$, the oriented priority \eqref{eq:orientedpriority} is of course the same as the movement capacity \eqref{eq:directed_capacity}.

The set $W_i$ denotes all output links that restrict the flow from $i$.\footnote{
	We defined the set $W_i$ that consisted of the output links restricting $i$ slightly differently in \citet[Definition 3.2]{wright_node_2017}
	Specifically, there we wrote the second condition as $\exists \; i' \neq i \text{ s.t. } p_{i',j^\ast} f_{i,j^\ast} \geq p_{i,j^\ast} f_{i',j^\ast}$.
	We have inverted the conditional in the present definition so that the definition is still valid for $M = 1$.}
The first condition for membership in $W_i$ ($\sum_c \beta_{i,j^\ast}^c S_i^c > 0$) can be read as ``there is some nonzero demand for the movement $i,j^\ast$.'' 

The second condition ($\nexists \; i' \neq i \text{ s.t. } \sfrac{f_{i,j^\ast}}{p_{i,j^\ast}} < \sfrac{f_{i',j^\ast}}{p_{i',j^\ast}}$) communicates that there is not \emph{some other} input link $i'$ that is able to send more (priority-normalized) flow to $j^\ast$: that is, that $j^\ast$ is either a) restricting to both $i$ and $i'$ (in which case we will have $\sfrac{f_{i,j^\ast}}{p_{i,j^\ast}} = \sfrac{f_{i',j^\ast}}{p_{i',j^\ast}}$ and $j^\ast \in W_i \cap W_{i'}$), or b) that $i'$'s flow to $j^\ast$ was demand-constrained.
If the opposite was true (that is, that $i'$ was able to send more (priority-normalized) flow to $j^\ast$ than $i$), then $j^\ast$ was not restricting after all (i.e., it was not the link that ran out of supply) and some other output link is what restricted the flow from $i$.
More discussion of a physical meaning for the second condition in \eqref{eq:scirw} will be given later, in Remark \ref{rem:scirw_t}.

Constraint \eqref{eq:scir1} says that if a link $i$ is not able to fill its demand, then there is at least one output link in $W_i$ that restricts $i$, and that $i$'s movements claim at least as much as their oriented-priority-proportional allocation of supply.
Constraint \eqref{eq:scir2} captures the reallocation of ``leftover'' supply, which states that a link $i$ that cannot fulfill all of its demand to the links in $W_i$ will continue to send vehicles after links $i': j \notin W_{i'}$ have fulfilled their demands to the $j \in W_i$.

\begin{rem}
	In \eqref{eq:orientedpriority}, we distribute the link priority among the movements proportionate to the demand for each movement.
	Consider the case where the link priority is chosen to be the link capacity, $p_i = F_i$, as we have mentioned as an example, and as suggested by \citet{tampere_generic_2011} (they argued the capacity makes sense as a priority value because vehicles leaving $i$ in a discharging queue will be claiming downstream supply as fast as possible, i.e., at the link's capacity).
	If the capacity is chosen based on a multiple of $i$'s number of lanes, note that this means that, in \eqref{eq:orientedpriority}, each movement will theoretically have available priority proportional to all lanes rather than only the lanes the movement can actually use.
	This will not be too unrealistic, however, if the demands $S_{i.j}$ are proportional to the number of lanes.
	A more refined SCIR where the oriented priorities $p_{i,j}$ are assigned in a manner aware of both the relative demands between the movements (as in \eqref{eq:orientedpriority}), and the spatial extent of the lane facilities the movements have available (similar to the partial FIFO construction), is an avenue for future work. 
\end{rem}

\subsubsection{Our first-order node model optimization problem}
Putting together the pieces, we have
\begin{defn}[First-order node model problem]
\begin{subequations} \label{eq:firstordergcnm}
\begin{equation}
\max\left(\sum_{i=1}^M\sum_{j=1}^N\sum_{c=1}^C f_{i,j}^c\right)
\label{eq:mimo_objective}
\end{equation}
subject to:
\begin{alignat}{2}
f_{i,j}^c &\geq 0 \;\; &\forall i,j,c
\label{mimo_nonnegativity_constraint} \\
f_{i,j}^c &\leq S_{i,j}^c &\forall i,j,c 
\label{mimo_demand_constraint} \\
\sum_i \sum_c f_{i,j}^c &\leq R_j, &\forall j
\label{mimo_supply_constraint} \\
\sum_i f_{i,j}^c &= \sum_j f_{i,j}^c &\forall c  \\
\frac{f_{i,j}^c}{\sum_c f^c_{i,j}} &= \frac{\beta_{i,j}^c S_{i}^c}{\sum_c \beta_{i,j}^c S^c_i} \;\; &\forall i,j,c \label{mimo_proportionality_constraint}
\end{alignat}
\begin{equation}
\textnormal{(Partial) FIFO and SCIR constraints (in this work, \eqref{eq:rectangleRelaxedFifo} and \eqref{eq:scir})}.
\end{equation}
\end{subequations}
\end{defn}

A solution will have flows that are constrained by at least one of the constraints outlined above.
An algorithm to solve this problem and proof of optimality was given in \citet{wright_node_2017}.
Below, we will present a new (simpler) algorithm for the same problem.

\subsubsection{Other first-order node model requirements}
\label{sec:other}
Note that the list of first-order node requirements presented so far \ref{sec:gcnmfirstorder} (which is the particular node problem of interest for the remainder of this work) is not an exhaustive list of all ``reasonable'' node model requirements.
Since the statement of the GCNM requirements in \citet{tampere_generic_2011}, several authors have proposed extensions or modifications (as we have in the ``partial FIFO'' relaxation).
Beyond what we have covered here, one of the most discussed are nodal supply constraints.
These supply constraints, as their name suggests, describe supply limitations at the node rather than in one of the output links.
They are meant to describe restrictions on traffic that occur due to interference between flows in the junction (rather than vehicles being blocked in the input link), or the exhaustion of some ``shared resource'' such as green light time at a signalized intersection.
Each movement through the node may or may not consume an amount of a node supply proportional to its throughflow.

The node supply constraints in the GCNM framework were originally proposed in \citet{tampere_generic_2011}.
In \citet{corthout_non-unique_2012} it was noted that these node supplies may lead to non-unique solutions.
\citet{jabari_node_2016} revisited the node supply constraints (mostly in the context of distribution of green time) to address \citet{corthout_non-unique_2012}'s critique of non-uniqueness of solutions and suggested that non-uniqueness can be resolved by properly accounting for signal phasing (i.e., which green time allocations are active at the same time).

We do not explicitly include the node supply constraints in the dynamic system node models and resulting solution algorithms in this work.
The path towards their inclusion in the first- and second-order cases is similar to the partial FIFO construction but notationally cumbersome and somewhat beyond this work's scope of fusing the GCNM and second-order link models.

\subsection{Other approaches to road junction modeling}
In this section so far, we mostly reviewed node models that pose the node flow problem as an optimization problem (e.g., \citet{tampere_generic_2011, flotterod_operational_2011, corthout_non-unique_2012,smits_family_2015, jabari_node_2016, wright_node_2017} and their references).
As mentioned in section \ref{gsom_node:sec:intro}, this type of problem setup can be interpreted as taking as input the adjacent links' supply and demand (i.e., their state at the boundary) and producing as output in- and out-flows (i.e., Neumann boundary conditions) for those links.
We will use this framework for the remainder of this work.

Beyond the solving-for-node-flows approach, another class of methods have seen recent development and should be mentioned.
As described in \citet{jin_riemann_2017}, these methods instead resolve the multidimensional Reimann problem by breaking it into one Riemann problem for each link, then solving each one.
Methods of this type tend to explicitly use the terminology of ``Riemann solvers'' (e.g., \citet{herty_coupling_2006,garavello_traffic_2006,haut_second_2007,jin_riemann_2017}) rather than ``node models.''
Compared to the node-flow framework, the Riemann-solver framework could be interpreted as setting up and solving a traditional PDE boundary value problem for each link.
One work of particular interest \emph{vis-\`{a}-vis} the node-flow framework is \citet{jin_riemann_2017}, which claims to bring these two classes of methods closer together by incorporating the concepts of supply and demand that the node-flow framework inherited from the one-dimensional discretization \citep{daganzo_cell_1995}.

More related to this work's concepts, though, are the works of \citet{herty_coupling_2006, garavello_traffic_2006, haut_second_2007}, which applied this second approach to junctions of roads with second-order dynamics.
These ``Riemann-solver-framework'' methods make a point to \emph{not} use the the terminology of demand $S$ and supply $R$.
As we will discuss in more detail in section \ref{sec:secondorder_discussion}, the particular differences of the second-order node flow problem relative to the first-order problem make the approach (as done in \eqref{sec:gcnmfirstorder}) of treating supply and demand as exogenous constraints not suitable to the second-order case.

The second-order generalization of the node problem, then, needs to take inspiration from the ``Riemann-solver-framework'' approach of not isolating the node flows from the surrounding links.
We will make this statement more concrete later.

\subsection{Review of first-order node dynamic system}
\label{sec:firstorderds}
This section reviews the node dynamic system characterization of node models presented in \citet{wright_dynamic_2016}.
This dynamic system is a hybrid system, which means that it contains both continuous and discrete states (also called discrete modes).
Here, the continuous states evolve in time according to differential equations, the differential equations themselves change between discrete states, and the discrete state transitions are activated when conditions on the continuous states are satisfied.

\begin{defn}[Generic first-order node hybrid system]
~\\
\begin{itemize}
	  \item Let there be $N \cdot M \cdot C$ time-varying continuous states $f_{i,j}^c(t)$, each representing the number of vehicles of commodity $c$ that have taken movement $i,j$ through the node.
	  The continuous state space is denoted $X$.
    \item Let $\mathcal{J}$ be the set of all output links $j$. Let there be $2^M$ discrete states $q_\nu, \nu \in 2^\mathcal{J}$ (recall $2^\mathcal{J}$ refers to the power set of $\mathcal{J}$), the index $\nu$ representing the set of downstream links that have become congested.
A downstream link $j$ is said to ``become congested'' at time $t$ if $\sum_i \sum_c f_{ij}^c(t) = R_j$.
The discrete state space is denoted $Q$.
    \item Init $\subseteq Q \times X$ defines the set of permissible initial states of the system at $t=0$.
    \item Dom: $Q \to X$ denotes the domain of a discrete state, which is the space of permissible continuous states while the discrete state is active.
    \item $\Phi: Q \times X \to Q \times X$ is a reset relation, which defines the transitions between discrete states and the conditions for those transitions.
		\item The hybrid system execution begins at time $t=0$ and discrete state index $k=0$.
		We will say that $k$ increments by one every time there is a transition between discrete states.
		\item Each link is given a ``time limit'' $T_i \triangleq \sfrac{F_i}{p_i}$.
		As we discussed in section \ref{sec:partialfifo}, the capacity $F_i$ is defined as the maximum possible flow of a given time interval, and as shown in \eqref{general:F} below, the maximum possible flow rate (when there is no partial FIFO blockage) for a link $i$ is the priority $p_i$, the partial-FIFO constraint is enforced in this manner.
	\end{itemize}
	Our hybrid system ($Q, X,$ Init, $\dot{f}_{i,j}^c$, Dom, $\Phi$) is
	\begin{subequations}
  \label{eq:firstorderHS}
	\begin{align}
		Q &= \{  q_\nu \}, \, \nu \in 2^\mathcal{J} \label{general:Q} \displaybreak[0]\\
		X &= \RR^{M \cdot N \cdot C} \label{general:X} \displaybreak[0]\\
		\textnormal{Init} &= Q \times \{f_{i,j}^c(t=0)=0 \, \quad \forall i,j,c\} \label{general:init} \displaybreak[0]\\
		\dot{f}_{i,j}^c(q,f) &= \begin{cases}
			  \begin{aligned}[c]
			  p_{i,j} \frac{ 
					S_{i,j}^c - f_{i,j}^c(t) }{
						\sum_c \left(S_{i,j}^c - f_{i,j}^c(t) \right)}
			  \bigg(1 - \big| \bigcup_{ \mathclap{
			  \substack{j' \in \nu(k),\\	\exists \, c: \, f_{i,j'}^c < S_{i,j'}^c }} }
			  \bm{\eta}_{j',j}^i \big| \bigg)
			  \end{aligned}
  			  & \text{ if } f_{i,j}^c(t) < S_{i,j}^c \text{ and } t < T_i \\
			  0 &\text{ otherwise}
		\end{cases}
		\label{general:F} \displaybreak[0]\\
			\textnormal{Dom}(q_\nu) &=
			\left\{ \begin{aligned} 
				f: &\sum_i \sum_c f_{i,j}^c = R_j \quad \hspace{-.25em} \forall j \in \nu \textnormal{ and} \\
					&\sum_i \sum_c f_{i,j}^c \leq R_j \quad \hspace{-.25em} \forall j \notin \nu
				\end{aligned} \right\} \label{general:dom} \displaybreak[0]\\
		\Phi(q_\nu,f) &= 
		\begin{aligned}[t]
			(q_{\nu'}, f) &\textnormal{ if } \sum_i \sum_c f_{i,j^\ast}^c = R_{j^\ast} \\
				&\textnormal{ where } \nu' = \nu \cup j^\ast.
		\end{aligned} \label{general:r}
	\end{align}
	\end{subequations}
	When $\dot{x}^c_{ij} = 0$ for all $i,j,c$, the execution is complete and the $f_{ij}^c$ take their final values.
\end{defn}

It was shown in \citet{wright_dynamic_2016} that the hybrid system \eqref{eq:firstorderHS} produces the same solutions as \citet{wright_node_2017}'s algorithm (again, for the prior version where the partial FIFO constraint was defined differently, which changed the dynamic system definition in the definition of the link time limit $T_i$).
In the following section, we show how to quickly compute executions of the hybrid system, which, since it is based on the continuous-time dynamics of \eqref{eq:firstorderHS}, presents a more intuitive algorithm than the one in \citet{wright_node_2017}.

\subsection{Execution of the first-order node dynamic system as a simple algorithm}
\label{sec:firstorder_algorithm}
Evaluating continuous-time or hybrid systems typically involves forward integration of the differential equation(s) with fixed or varying step sizes.
However, in the case of \eqref{eq:firstorderHS}, evaluation can be performed in a much simpler manner.
This is due to the particular dynamics of the system - since the continuous-time dynamics and the condition for discrete mode switching are very simple, the time that the next mode switch will occur can be found in closed form.
Equations \eqref{general:dom} and \eqref{general:r} say that a mode switch where link $j$ enters $\nu$ will occur when
\begin{equation}
	\sum_i \sum_c f_{ij}^c = R_{j}. \label{eq:modeSwitch1}
\end{equation}
Say we are currently at time $t_0$. 
Combining \eqref{eq:modeSwitch1} with \eqref{general:F}, we can find the time that the mode switch occurs, which we denote $t_{j}$.
\begin{equation}
	R_{j} = \sum_i \sum_c f_{ij}^c(t_0) + \int_{t_0}^{t_j} \sum_i \sum_c \dot{f}_{ij}^c dt. \label{eq:modeSwitch2}
\end{equation}
Solving the integral in \eqref{eq:modeSwitch2},
\begin{align}
	\int_{t_0}^{t_j} \sum_i \sum_c \dot{f}_{i,j}^c dt
		&= \int_{t_0}^{t_j} \sum_i \sum_c p_{i,j} 
		\frac{ S_{i,j}^c - f_{i,j}^c(t) }{
						\sum_c \left(S_{i,j}^c - f_{i,j}^c(t) \right)}
	\bigg(1 - \big| \bigcup_{ \mathclap{
			\substack{j' \in \nu,\\	\exists \, c: \, x_{i,j'}^c < S_{i,j'}^c }} }
			\bm{\eta}_{j',j}^i \big| \bigg) dt
	\nonumber \displaybreak[0]\\
	&= \int_{t_0}^{t_j} \sum_i p_{i,j}
	\bigg(1 - \big| \bigcup_{ \mathclap{
			\substack{j' \in \nu,\\	\exists \, c: \, f_{i,j'}^c < S_{i,j'}^c }} }
			\bm{\eta}_{j',j}^i \big| \bigg) \nonumber \displaybreak[0]\\
	&= (t_j - t_0) \sum_i p_{i,j}
		\bigg(1 - \big| \bigcup_{ \mathclap{
				\substack{j' \in \nu,\\	\exists \, c: \, f_{i,j'}^c < S_{i,j'}^c }} }
				\bm{\eta}_{j',j}^i \big| \bigg). \label{eq:modeSwitch3}
\end{align}
Then, plugging \eqref{eq:modeSwitch3} into \eqref{eq:modeSwitch2},
\begin{equation}
	t_j = t_0 + \frac{ R_j - \sum_i \sum_c f_{i,j}^c(t_0)}{ \sum_i p_{i,j} \bigg( 1 - \big| \bigcup\limits_{ \mathclap{
					\substack{j' \in \nu,\\	\exists \, c: \, f_{i,j'}^c < S_{i,j'}^c }} }
					 \bm{\eta}_{j',j}^i \big| \bigg) }. \label{eq:modeSwitch4}
\end{equation}
This value can be computed for each output link $j$.
Then, the $j$ with the smallest $t_j$ will be the first link to fill and join $\nu$.
We had used $j^\ast$ for this output link, so let $t_{j^\ast} \triangleq \min t_j$.
However, one of the input links may have its time limit $T_i$ expire.
This would also change the dynamics, as it stops sending vehicles at that time.

Therefore, evaluation of the system trajectory beginning from $t_0$ can be done by (i) evaluating \eqref{eq:modeSwitch4} for each output link, (ii) identifying $t_j^\ast$, and (iii) checking whether any of the time limits $T_i$ occur before $t_{j^\ast}$.
This is an event-triggered simulation: it is only necessary to determine when the next event will occur.
The equations for $\dot{x}_{ij}^c$ over $[t_0, \min (\{T_i\}, t_{j^\ast})]$ can then be evaluated in closed form under $q_\nu$.

Note that the $\dot{x}_{i,j}^c$'s for an $i$ may change to zero from nonzero without a change in the discrete state $q_\nu$, if the conditional of $x_{i,j}^c(t) < S_{i,j}^c$ in \eqref{general:F} is broken.
This can be understood as the $i$ running out of vehicles that it is able to send.
This may happen if $p_i > S_i$ for that $i$, and some (partial) FIFO constraint becomes active on $i$.
In the following algorithm, we introduce a new set, $\mu$, that was not present in the dynamic system definition and contains the $i$'s that either exhaust their supply or have their time limits expire (i.e., those $i$'s whose $\dot{x}_{i,j}^c$ become zero without $j$ necessarily entering $\nu$).

These steps are summarized in the Algorithm below.
This algorithm represents the completion of an argument began in \citet{wright_dynamic_2016}.

\begin{defn}[First-order node model solution algorithm] \label{def:1o_alg}
~\\

\textbf{List of inputs:}
	\begin{itemize}
		\item Per-commodity input demands $S_i^c \; \forall \; i,c$
		\item Per-output supplies $R_j \; \forall \; j$
		\item Per-commodity split ratios $\beta_{i,j}^c \; \forall \; i,j,c$
		\item Per-input priorities $p_i \; \forall \; i$
		\item Per-movement restriction intervals $\eta_{j',j}^i \; \forall \; i,j,j'$
	\end{itemize}
\begin{enumerate}
	\item \textbf{Initialize.}
					\begin{align*}
			k &= 0 & \text{Set iteration counter to 0} \\
			T_i &= \frac{\sum_c S_i^c}{p_i} & \text{Compute input ``time limits''} \\
			S_{i,j}^c &= \beta_{i,j}^c S_i & \text{Compute directed demands} \\
			p_{i,j} &= p_i \frac{\sum_c S_{i,j}^c}{\sum_c S_i^c} & \text{Compute oriented priorities} \\
			\nu(k=0) &= \{j: R_j = 0 \} & \text{Note initial $(k=0)$ zero-supply output links} \\
			f_{i,j}^c(k=0) &= 0 & \text{Initialize all throughflows to 0} \\
		\end{align*}
												\item \textbf{Compute flow rates.} \label{step:flowrate}
			\begin{align}
		\dot{f}_{i,j}^c(k) = \begin{cases}
			  \begin{aligned}[c]
			  p_{i,j} 
						  \frac{ 
				S_{i,j}^c - f_{i,j}^c(t) }{
					\sum_c \left(S_{i,j}^c - f_{i,j}^c(t) \right)}
			  \bigg(1 - \big| \bigcup_{ \mathclap{
			  \substack{j' \in \nu(k),\\	\exists \, c: \, f_{i,j'}^c < S_{i,j'}^c }} }
			  \bm{\eta}_{j',j}^i \big| \bigg)
			  \end{aligned}
  			  & \text{ if } f_{i,j}^c(k) < S_{i,j}^c \text{ and } t(k) < T_i \\
			  0 &\text{ otherwise}
		  \end{cases}
		  \quad \forall \; i,j,c \label{eq:flowrate}
			\end{align}
		\item \textbf{If} $\dot{f}_{i,j}^c(k) = 0 \; \forall \; i,j,c$, \textbf{then the algorithm ends. Otherwise, continue.}

	\item \textbf{Compute the length of this iteration's timestep.} \label{item:firstorder_timestep}
	\begin{align}
		t_j(k) &= \frac{R_j - \sum_i \sum_c f_{i,j}^c(k)}{\sum_c \sum_i \dot{f}_{i,j}^c(k)}  & \forall \; j \nonumber \\
							dt(k) &= \min\left\{
			 			 \{t_j(k) - t(k) \}_{j \notin \nu(k)}, \;
			 \{T_i - t(k)\}_{i \notin \mu(k)} \right\} & \nonumber
	\end{align}
	\item \textbf{Advance forward in time.}
	\begin{align*}
		\Delta f_{i,j}^c(k) &= \dot{f}_{i,j}^c(k) \cdot dt(k) & \forall \; i,j,c \\
		f_{i,j}^c(k+1) &= f_{i,j}^c(k) + \Delta f_{i,j}^c(k) & \forall \; i,j,c \\
		t(k+1) &= t(k) + dt(k) &
	\end{align*}
	\item \textbf{Update set of ``completed'' links.}
	\begin{align*}
		\nu(k+1) &= \nu(k) \cup \left\{j: R_j - \sum_i \sum_c f_{i,j}^c(k+1) = 0 \right\}
	\end{align*}
	\item \textbf{Set} $k \gets k+1$, \textbf{return to step \ref{step:flowrate} and repeat.}
\end{enumerate}

\end{defn}

We consider the algorithm given in Definition \ref{def:1o_alg} to be quite a bit more intuitive than previous forms of solution algorithms to the node model problem that do not follow the dynamic system approach.
Key to this is that all the quantities have a clear physical meaning.

\begin{rem} \label{rem:scirw_t}
Now that we have introduced the concept of continuous-time timesteps in this context (Step \ref{item:firstorder_timestep} above), we can briefly revisit our specific SCIR \eqref{eq:scir} to provide a physical intuition for the second condition in \eqref{eq:scirw}.
Recalling the second condition in \eqref{eq:scirw},
\begin{equation*}
	\frac{f_{i, j^\ast}}{p_{i, j^\ast}} < \frac{f_{i', j^\ast}}{p_{i', j^\ast}}
\end{equation*}
and recalling \eqref{eq:flowrate}, an un-restricted flow from $i$ to $j$ will have flow rate $\dot{f}_{i,j}^c = p_i (\sfrac{S_{i,j}^c}{( \sum_c S_{i,j}^c)})$.
So, \eqref{eq:scirw} can be rewritten as
\begin{equation*}
	\frac{f_{i, j^\ast}}{\dot{f}_{i, j^\ast}} < \frac{f_{i', j^\ast}}{\dot{f}_{i', j^\ast}}.
\end{equation*}
In other words, the condition that states that $i$ cannot be restricted by $j^\ast$ (and must be restricted by some other constraint) if another link $i'$ is able to send more priority-normalized flow to $j^\ast$ also means that $i'$ cannot send flow to $j^\ast$ for a \emph{longer time} than $i$.
\end{rem}

\section{Review of second-order flow modeling}
\label{sec:secondorder}
\subsection{Introduction}
\label{sec:secondorder_intro}
The formulation of the GSOM seen in \eqref{eq:gsom} has been called the ``advective form'' \citep{fan_collapsed_2017}.
In this form, the property $w$ is advected with the vehicles at speed $v$.
That is, it is constant along trajectories.
This form makes the statement that the property $w$ is a property of vehicles.
This understanding is useful because it communicates how oftentimes the forms of $w$ and $V(\rho,w)$ are constructed such that they model some microscopic, per-driver behavior.

For example, it has been shown \citep{zhang_non-equilibrium_2002,aw_derivation_2002} that the original ARZ model can be characterized as a ``coarsening'' to a macro-scale of simple car-following models of the form
\begin{align}
	\begin{split}
	\ddot{x}_n(t) &= \frac{\dot{x}_{n-1}(t) - \dot{x}_n(t)}{\tau(s_n(t)) } \\
	s_n(t) &= x_{n-1}(t) - x_n(t) \label{eq:carfollowing}
	\end{split}
\end{align}
where $x_n(t)$, $\dot{x}_n(t)$ and $\ddot{x}_n(t)$ are the position, velocity, and acceleration of the $n$th car, respectively, and $\tau(s_n(t))$ is a driver's response time, which is stated to be a function of their distance from the car ahead of them.
This equivalence is shown by setting the velocity function as $V(\rho,w) = V_{\textnormal{eq}}(\rho) + (w - V_{\textnormal{eq}}(0))$, with $V_{\textnormal{eq}}$ defined as an \emph{equilibrium} velocity function.
The advected property $w$ is thought of as the driver's distance from equilibrium velocity \citep{aw_derivation_2002, zhang_non-equilibrium_2002, lebacque_generic_2007}.\footnote{
	Of potential interest to the reader may be the analysis of \citet{aw_derivation_2002}, where a particular form of $\tau(\cdot)$ is shown to be equivalent to a form of \eqref{eq:gsom} with an inhomogeneous \eqref{eq:gsom2}: that is, with $w$ allowed to be created or destroyed.
	We do not consider inhomogeneous forms of \eqref{eq:gsom} in this work.
}

Each individual vehicle can be thought of as having its own $w$, and the macroscopic $w$ in the PDE form \eqref{eq:gsom2} then equals the average of the vehicle $w$'s.
To apply a discretization to this PDE formulation, it is useful to consider the \emph{total property} $\rho w$, and rewrite \eqref{eq:gsom} in ``conservative form'' \citep{lebacque_generic_2007, fan_collapsed_2017},
\begin{align}
\begin{split}
  \frac{\partial \rho}{ \partial t} + \frac{ \partial (\rho v)}{\partial z} &= 0 \\
  \frac{\partial (\rho w)}{ \partial t} + \frac{ \partial (\rho w v)}{\partial z} &= 0 \\
  \textnormal{where } v &= V(\rho, w). \label{eq:gsom_conservative}
\end{split}
\end{align}
We will review the relevant finite-volume discretization using the Godunov scheme of \eqref{eq:gsom_conservative} \citep{lebacque_generic_2007} in the next section.
For a deeper analysis on the physical properties of \eqref{eq:gsom_conservative}, see, e.g., \citet{lebacque_generic_2007}.

We make one note on constraints imposed on the form of $v(\rho, w)$ in \eqref{eq:gsom_conservative}.
It has been stated \citep[(19)]{lebacque_generic_2007} that, to apply the Godunov discretization to \eqref{eq:gsom_conservative}, one is restricted to choices of $V(\rho,w)$ for which there is a unique $\rho$ for every $(v,w), v \neq 0$ and a unique $w$ for every $(v,\rho), v \neq 0$.
That is, $V(\rho,w)$ must be invertible in both its arguments for nonzero values of velocity.

\begin{rem}
In this work, the only further assumption we make on the form of $V(\cdot)$ is that $v=0$ occurs only for a specific maximum value of density, $\rho_{max}$, and that at that density, $V(\rho_{max}, w) = 0$ for any value of $w$.
Further, we assume that the converse is not true: that no value of $w$ exists that induces $v=0$ for $\rho < \rho_{max}$.
\end{rem}

\subsection{Godunov discretization of the GSOM}
\label{sec:2o_onetoone}
The Godunov discretization of the first-order (LWR) model \eqref{prelim:eq:LWR}, first introduced as the Cell Transmission Model \citep{daganzo_cell_1994} is well-known.
The Godunov scheme discretizes a conservation law into small finite-volume cells.
Each cell has a constant value of the conserved quantity, and inter-cell fluxes are computed by solving Riemann problems at each boundary.
The solution to each Riemann problem is a flux that describes the amount of the conserved quantity that is sent from the upstream cell to the downstream cell.
The Godunov scheme is a first-order method, so it is useful for simulating solutions to PDEs with no second- or higher-order derivatives like the LWR formulation.
In the CTM, the Godunov flux problem is stated in the form of the demand and supply functions.

Since \eqref{eq:gsom_conservative} is also a conservation law with no second- or higher-order derivatives, the Godunov scheme is applicable as well \citep{lebacque_generic_2007}.
However, due to the second PDE for $\rho w$, an \emph{intermediate state} arises in the Riemann problem and its solution \citep{aw_resurrection_2000, zhang_non-equilibrium_2002, lebacque_awrascle_2007, lebacque_generic_2007}.
This intermediate state has not always had a clear physical meaning, and this lack of clarity likely inhibited the extension of the Godunov discretization to the multi-input multi-output node case.
In our following outline of the discretized one-input one-output flow problem, we make use of a physical interpretation of the intermediate state due to \citet{fan_collapsed_2017}.

A final note: in the first-order node model, we were able to ignore the first-order demand and supply functions that generated the supplies $R_j$ and per-commodities demands $S_i^c$.
That is, we were agnostic to the method by which they were computed (and to the input and output link densities), as they did not change during evaluation of the node problem.
As we will see shortly, this is not the case for the second-order flow problem (due to the intermediate state and its interactions with the downstream link).
Therefore, our explanation below makes use of the second-order demand and supply functions $S(\rho, w)$ and $R(\rho, w)$, respectively.

\subsubsection{Preliminaries}
\label{sec:second_prelims}

In this work, we say that each vehicle commodity $c$ has its own property value $w^c$.
The net (averaged over vehicle commodities) property of a link $\ell$, denoted $w_\ell$, is
\begin{equation}
  w_\ell = \frac{\sum_c w^c \rho_\ell^c}{\rho_\ell} \label{eq:link_w}
\end{equation}
where $\rho_\ell \triangleq \sum_c \rho_\ell^c$ is the total density of link $\ell$.

In the second-order model, the fundamental diagram of a link is a function of both net density and net property as defined above.
This carries over to the demand and supply functions in the Godunov discretization \citep{lebacque_generic_2007, fan_collapsed_2017}.
That means that the supply and demand are defined at the link level with the net quantities $\rho_\ell$ and $w_\ell$.
For an input link $i$,
\begin{equation}
  S_i = S_i( \rho_i, w_i) =
  \begin{cases}
    \rho_i v_i & \textnormal{if } \rho_i \leq \rho_c (w_i) \\
    F(w_i) & \textnormal{if } \rho_i > \rho_c (w_i)
    \end{cases} \label{eq:demanddefined}
\end{equation}
where $\rho_c(w_i)$ is the critical density for property value $w_i$ and $F(w_i)$ is the capacity for property value $w_i$.

The demand from \eqref{eq:demanddefined} is split among the commodities and movements proportional to their densities and split ratios,
\begin{align*}
  S_i^c &= S_i \frac{\rho_i^c}{\rho_i} \\
  S_{i,j}^c &= \beta_{i,j}^c S_i^c.
\end{align*}

The oriented priorities $p_{i,j}$ are computed according to \eqref{eq:orientedpriority}, as before.

\subsubsection{Computing supply}
\label{sec:2o_supply}
Compared to computing the demand, solving for an output link's supply in the multi-input-multi-output second-order case is a much more complicated problem.
We will begin our discussion with a review of the one-input-one-output case \citep[Sections 3.3, 3.4]{fan_collapsed_2017}.

The supply $R$ of the output link in this one-to-one case, where $i$ is the input link and $j$ is the output link, is
\begin{equation}
  R_j = R_j(\rho_M, w_i) =
  \begin{cases}
    F(w_i) & \textnormal{if } \rho_M \leq \rho_c(w_i) \\
    \rho_M v_M & \textnormal{if }\rho_M > \rho_c(w_i).
  \end{cases} \label{eq:1to1_R}
\end{equation}
We see that the supply of the downstream link is actually a function of the upstream link's vehicles' property, and the density and speed of some ``middle'' state, $M$.
The middle state is given by \cite[(16)]{fan_collapsed_2017}
\begin{subequations}
\label{eq:middle_state}
\begin{align}
  w_M &= w_i \label{eq:middle_w}\\
  v_M &= \begin{cases}
    V(0, w_i) & \textnormal{if } V(0, w_i) < v_j \\
    v_j & \textnormal{otherwise}
    \end{cases} \label{eq:middle_v} \\
  \rho_M &\textnormal{ s.t. } v_M = V( \rho_M, w_M ) \label{eq:middle_rho}
\end{align}
\end{subequations}
where $v_j = V(\rho_j, w_j)$ is the velocity of the downstream link's vehicles and $V(\cdot)$ is the velocity function as given by the fundamental diagram.
It is assumed that 

It is worth emphasizing that the speed of the flow between two links, $v_M$, is equal across all vehicles that are moving, despite each individual vehicle possibly having different $w$'s.
This is, again, due to the isotropic commodity mixing assumption.
The particular value at which the flow moves is dependent on the net $w$ of this isotropic mixture \eqref{eq:link_w}, as shown in \eqref{eq:middle_state}.

In \citet{fan_collapsed_2017}, the intuition behind the meaning of the middle state is given as follows: the middle state vehicles are actually those that are leaving the upstream link $i$ and entering the downstream link $j$.
As they leave $i$ and enter $j$, they clearly carry their own property \eqref{eq:middle_w}, but their velocity is upper-bounded by the velocity at which that the downstream vehicles exit link $j$ and free up the space that the $i$-to-$j$ vehicles enter \eqref{eq:middle_v}.
The middle density, $\rho_M$ (and therefore the downstream supply $R$), is then determined by both the upstream vehicles' characteristics (i.e., $w_i$) and the downstream link's flow characteristics (through $v_j$).
In other words, the number of vehicles that can fit into whatever space is freed up in the downstream link is a function of the drivers' willingness to pack together (defined by $w_i$).
Since the meaning of supply $R_j$ is ``the number of vehicles that $j$ can accept,'' this means that $R_j$ is dependent on $w_i$ \eqref{eq:middle_rho}.

Note that \eqref{eq:middle_v} is also the equation by which congestion spills back from $j$ to $i$: if $j$ is highly congested, then $v_j$ will be low.
This then makes $\rho_M$ large in \eqref{eq:middle_rho}, which in turn leads to a small $R_j$ in \eqref{eq:1to1_R}. 

Now that we have reviewed the 1-to-1 case, we can consider how to generalize this to a multi-input-multi-output node when we determine supply for several links.

\section{Second-order node model problem and solution}
\label{sec:secondordernode}
\subsection{Multi-input-multi-output extension of the second-order Godunov discretization}
\label{sec:2omimo}
We saw that the reasoning behind the dependence of $R_j$ on $w_i$ was that the spacing tendencies of $i$'s vehicles determine the number of vehicles that can fit in $j$.
Therefore, in generalizing to a multi-input-multi-output node, it makes sense to define a link $j$'s ``middle state'' as being dependent on the vehicles \emph{actually entering} link $j$.
That is, if $w_{j^-}$, the $w$ just upstream of $j$, is the ``middle state'' of link $j$, then we say
\begin{subequations}
\begin{equation}
  w_{j^-} = \frac{\sum_i \sum_c w^c \dot{f}_{i,j}^c }{\sum_i \sum_c \dot{f}_{i,j}^c}. \label{eq:mimo_middlew}
\end{equation}
The ``$j$-upstream middle state'' velocity and density, $v_{j^-}$ and $\rho_{j^-}$, are then
\begin{align}
  v_{j^-} &= \begin{cases}
    V(0, w_{j^-}) & \textnormal{if } V(0, w_{j^-}) < v_j \\
    v_j & \textnormal{otherwise}
    \end{cases} \label{eq:mimo_middlev} \\
  \rho_{j^-} &\textnormal{ s.t. } v_{j^-} = V( \rho_{j^-}, w_{j^-} )
\end{align} \label{eq:mimo_middle}
\end{subequations}
and the supply $R_j$ is
\begin{equation}
  R_j = R_j(\rho_{j^-}, w_{j^-}) =
  \begin{cases}
    F(w_{j^-}) & \textnormal{if } \rho_{j^-} \leq \rho_c(w_{j^-}) \\
    \rho_{j^-} v_{j^-} & \textnormal{if }\rho_{j^-} > \rho_c(w_{j^-}).
  \end{cases} \label{eq:mimoR}
\end{equation}

Note that in \eqref{eq:mimo_middlew}, we defined $w_{j^-}$ as a function of $\dot{f}_{ij}^c$.
Recall from the first-order node model that the $\dot{f}_{ij}^c$'s can change as (i) upstream links $i$ exhaust their demand or (ii) downstream links $j$ run out of supply.
These two events correspond to discrete state changes in our hybrid system.
This, of course, carries over to the second-order node model.
This means that the $j^-$ quantities, and thus the supply $R_j$, change as $\dot{f}_{ij}^c$'s change.
Therefore, at each discrete state transition, we need to determine the new supply for each output link $j$ for the new mixture of vehicles that will be entering $j$ in the next discrete state.

We will explain how this is done through the following example.
Suppose that at time $t_0$, we compute some $w_{j^-}, v_{j^-}, \rho_{j^-}$, and $R_j$ with \eqref{eq:mimo_middle}-\eqref{eq:mimoR}.
Then, at time $t_1$, one of the $\dot{f}_{i,j}^c$ for that $j$ changes.
At that point, we recompute $\rho_j$ and $w_j$,
\begin{align}
  \begin{split}
  \rho_j^c(t_1) &= \rho_j^c(t_0) + \frac{1}{L_j} \sum_i f_{ij}^c(t_1) \\
  \rho_j(t_1) &= \sum_c \rho_j^c(t_1) \\
  w_j(t_1) &= \frac{\sum_c w^c \rho_j^c(t_1)}{\rho_j(t_1)}
  \end{split} \label{eq:recompute_j}
\end{align}
where $L_j$ is the length of $j$.
Then, we recompute all the ``middle state'' variables and $R_j$ using \eqref{eq:mimo_middle}-\eqref{eq:mimoR}.
Critically, note that in this recomputation, the new $v_j$ at $t_1$ is $v_j(t_1) = V(\rho_j(t_1), w_j(t_1))$.
This means that $v_{j^-}(t_1)$ will also be different than $v_{j^-}(t_0)$.
This will carry through to create a $R_j(t_1)$ that is different from $R_j(t_0)$, and takes into account both the vehicles that have moved into $j$ between $t_0$ and $t_1$, and the difference in properties $w_{j^-}(t_0)$ and $w_{j^-}(t_1)$.

Note that if $w_{j^-}(t_1)$ leads to significantly tighter packing (i.e., smaller inter-vehicle spacing) than $w_{j^-}(t_0)$, it is conceivable that we will have $R_j(t_1) > R_j(t_0)$ (especially if $\rho_j(t_0)$ is not that much smaller than $\rho_j(t_1)$).
\footnote{As an illustrative example, suppose that we had an output link $j$ with two input links.
One input link has high priority and emits large trucks that have high inter-vehicle spacing, while the other input link has low priority and emits low-spacing cars.
Suppose at time $t_0$ both input links are sending vehicles, which would lead to a $w_{j^-}(t_0)$ that is weighted towards the (high-priority) trucks.
But then suppose the high-priority input link exhausts its demand of trucks quickly (relative to $R_j(t_0)$).
If we call this time of exhaustion $t_1$, then we will have a new $w_{j^-}(t_1)$ that reflects the low-spacing cars.
It is possible that in this situation $R_j(t_1) > R_j(t_0)$, since the units of $R_j(t)$ are in \emph{vehicles}, which is weighted by the \emph{type} of vehicles entering:
at time $t_1$, a slightly smaller amount of road is available than at time $t_0$, but since the mixture of vehicles at $t_1$ are themselves are smaller and will pack tighter, we can fit more into that amount of road.
}

Of course, the description above assumes isotropic mixing of all vehicle commodities in the link $j$ (recall we stated this assumption for input links $i$ in item \eqref{item:classprop} of the first-order GCNM requirements in section \ref{sec:gcnmfirstorder}).

Unlike supply, demand does not need to be recomputed since we assume the mixture of vehicles \emph{demanding} each movement remains the same (due to our isotropic mixture assumption)

In summary, the second-order multi-input-multi-output case is distinct from both the first-order multi-input-multi-output case and the second-order single-input-single-output case in that we cannot define a single supply $R_j$ at the time of problem statement.
Instead, the eventual $R_j$, the maximum amount of vehicles that $j$ can accept, will be dependent on $w_{j^-}$, which is itself dependent on the $f_{i,j}^c$'s for that $j$.
This circular dependency greatly exacerbates the complexity of the node problem, as we will see in the next section.

\subsection{Generic class of second-order node models: Problem Statement}
\label{sec:secondorderstatement}
We now present our proposed statement of the second-order node problem.
\begin{defn}[Generic second-order node model problem]
\begin{subequations} \label{eq:secondordergcnm}
\begin{equation}
\max\left(\sum_{i=1}^M\sum_{j=1}^N\sum_{c=1}^C f_{i,j}^c\right)
\label{eq:secondorder_objective}
\end{equation}
subject to:
\begin{alignat}{2}
f_{i,j}^c &\geq 0 \;\; &\forall i,j,c
\label{secondorder_nonnegativity_constraint} \\
f_{i,j}^c &\leq S_{i,j}^c \;\; &\forall i,j,c 
\label{secondorder_demand_constraint} \\
\sum_{i=1}^M \sum_{c=1}^C f_{i,j}^c &\leq R_j(\rho_{j^-}, w_{j^{-}}) &\forall j
\label{secondroder_supply_constraint} \\
\sum_i f_{i,j}^c &= \sum_j f_{i,j}^c &\forall c  \\
\frac{f_{i,j}^c}{\sum_c f^c_{i,j}} &= \frac{ S_{i,j}^c}{\sum_c S^c_{i,j}} \;\; &\forall i,j,c \label{secondroder_proportionality_constraint}
\end{alignat}
\begin{equation}
\textnormal{(Partial) FIFO and SCIR constraints \label{eq:gcnm_secondorder_fifoscir}
  }
\end{equation}
where $R_j(\rho_{j^-}, w_{j^-})$ is given by \eqref{eq:mimoR}, and where, as before,
\begin{align*}
	w_i &= \frac{\sum_c w^c \rho_i^c}{\rho_i} \\
	\rho_i &= \sum_c \rho_i^c \\
	S_i^c &= S_i(\rho_i, w_i) \frac{\rho_i^c}{\rho_i} \label{} \\
	S_{i,j}^c &= \beta_{i,j}^c S_i^c
\end{align*}
and where the ``$j$-upstream middle state'' is a modified version of \eqref{eq:mimo_middle} in that \eqref{eq:mimo_middlew} is integrated over time,
\begin{alignat}{2}
	w_{j^-} &= \frac{\sum_i \sum_c w^c f_{i,j}^c}{\sum_i \sum_c f_{i,j}^c} &\forall j \label{eq:node_upstream_w}\\
	  v_{j^-} &= \begin{cases}
    V(0, w_{j^-}) & \textnormal{if } V(0, w_{j^-}) < v_j \\
    v_j & \textnormal{otherwise}
    \end{cases} \;\; & \forall j \\
  \rho_{j^-} &\textnormal{ s.t. } v_{j^-} = V( \rho_{j^-}, w_{j^-} ) &\forall j \label{eq:node_upstream_rho}
\end{alignat}
\end{subequations}
\end{defn}
\begin{rem}
	Conservation of $w$ is enforced through \eqref{eq:node_upstream_w}.
\end{rem}
\begin{rem}
	Examining \eqref{secondroder_proportionality_constraint} and the definition of $S_{i,j}^c$, we see that, again, following the isotropic mixing assumption, all vehicles in a flow $(i.j)$ are equally impeded if the movement's downstream link $j$ is congested, even if individual commodities have drastically different $w$'s.
	The way that $w$'s affect the flow $(i,j)$ is through their contribution to the ``$j-upstream$'' state in \eqref{eq:node_upstream_w}.
	That is, in the macroscopic model, we do not model effects of $w$ variability (e.g., in-movement overtaking) \emph{within} an $(i,j)$ flow.
\end{rem}
The functions $S_i(\rho, w)$ and $R_j(\rho, w)$ are the particular demand and supply functions for those links.

In section \ref{sec:2omimo}, we noted that a circular dependency exists between the supplies $R_j$ and the flows $f_{i,j}^c$.
The form that this dependency takes in the problem statement above is in \eqref{secondroder_supply_constraint} and \eqref{eq:node_upstream_w}-\eqref{eq:node_upstream_rho}.
Plugging the definitions of $w_{j^-}$ and $\rho_{j^-}$ into \eqref{secondroder_supply_constraint}, we see that the $f_{i,j}^c$'s are constrained by another quantity that is a nonlinear function (through the fundamental diagram) of $f_{i,j}^c$.
We conjecture that this makes the optimization problem \eqref{eq:secondordergcnm} nonconvex for general fundamental diagrams.
In addition, this construction makes clear that the downstream supplies in the second-order multi-input-multi-output cannot be stated \emph{a priori}.
We discuss some implications of this fact in the next section.

\subsection{Generic Class of Second-order Node Models problem: Discussion}
\label{sec:secondorder_discussion}
In the first-order case as reviewed in section \ref{sec:firstorder}, the node problem is usually stated in terms of supply and demand instead of the actual conserved quantity $\rho$.
This is done to separate the node and link models from each other: once the link model(s) have been used to produce demand and supply, the node model problem is decoupled.
In addition, it is possibly more intuitive to think of flows as a function of supplies and demands (which have the same units as flow), rather than as a function of density discontinuities (in other words, we can abstract away the Riemann problem in the first-order case).

However, we have seen that a construction that abstracts away the link states in the node model is not possible for the general second-order case.
This is because the downstream supplies are dependent on the upstream $w_i$'s of the flows that enter each downstream link.\footnote{In the case of the one-to-one junction described in section \ref{sec:2o_onetoone}, we \emph{could} abstract away the supplies' dependency on the upstream $w_i$ and find an \emph{a priori} statement of the downstream $R_j$, since there was only one upstream $w_i$ to influence the supply in \eqref{eq:mimo_middlew}}

In our view, this difficulty in applying the extremely-useful ``supply-and-demand-based'' node problem setup to the second-order setting has likely hampered efforts towards defining and solving this extension.
Compare, for example, the several papers that have made progress towards the second-order junction problem in a ``traditional'' PDE, non-Godunov-based approach (e.g., \citep{herty_coupling_2006,garavello_traffic_2006,haut_second_2007}), that explicitly do not decouple the junction flow problem from the PDE model on links via the Godunov supply and demand functions.

In sum, when discussing the node model problem in the second order, one should take care to not think of a downstream link's ``supply'' in an unqualified sense.
Rather, the downstream link has a supply $R_j$ for each pair $(\rho_{j^-}, w_{j^-})$ \eqref{secondroder_supply_constraint}, one of which is compatible with the solution of the optimization problem.

However, our dynamic system construction for the node model solution generalizes to this case rather well.
This is described next.

\subsection{Dynamic system definition}
\label{sec:secondorderds}
Since we have defined the downstream supplies $R_j$ as depending on the net property of the flows \emph{actually entering} link $j$ \eqref{eq:mimo_middle}--\eqref{eq:mimoR}, and the final $R_j$ of the node problem solution \eqref{eq:secondordergcnm} as depending on the integrated-over-time flows into link $j$ \eqref{eq:node_upstream_w}, all we have to do to generalize the first-order node model dynamic construction to the second-order case is add a recomputation of the downstream supplies $R_j$ whenever we also recompute the $\dot{x}_{i,j}^c$'s due to a discrete mode switch.

In this dynamic system model and solution algorithm, we use as the partial FIFO and SCIR constraints referred in \eqref{eq:gcnm_secondorder_fifoscir} second-order generalizations of \eqref{eq:rectangleRelaxedFifo} and \eqref{eq:scir} such that supply is now a function of the ``j-upstream middle state.''
Specifically,
	\begin{equation}
	f_{i,j}^c \leq F_{i,j} \left(1 - \mathcal{A} \left( \bigcup_{j' \neq j} 
		\left\{ 
			\bm{\eta}_{j',j}^i \! \times \! \left[ \frac{f_{i,j'}}{F_{i,j'}} , \, 1 \right] \right\} \right)
		\right) \label{eq:rectangleRelaxedFifo_2o}
\end{equation}
and
\begin{subequations} \label{eq:scir_secondorder}
\begin{align}
	& \sum_j f_{i,j} < \sum_c S_i^c \implies W_i \neq \emptyset \quad \forall \, i \label{eq:scir1_secondorder} \displaybreak[0] \\
	& f_{i,j} \geq \frac{
	p_{i,j}}{  \sum_{i' = 1}^M p_{i',j}} R_j(\rho_{j^-}, w_{j^-}) \quad \forall \, j \in W_i \quad
	\forall i \label{eq:scir2_secondorder}
\end{align}
where 
\begin{equation}
	W_i = \left\{ j^\ast : \;
		\sum_c \beta_{i,j^\ast}^c S_i^c > 0 \text{ and } \nexists \; i' \neq i \text{ s.t. } \frac{f_{i,j^\ast}}{p_{i,j^\ast}} < \frac{f_{i',j^\ast}}{p_{i',j^\ast}}\right\} \label{eq:scirw_secondorder}
\end{equation}
\end{subequations}
and 
\begin{subequations}
\begin{align}
  p_{i,j} &= \frac{\sum_c S_{i,j}^c}{\sum_c S_i^c} p_i \label{eq:orientedpriority_secondorder} \\
  F_{i,j} &= \frac{\sum_c S_{i,j}^c}{\sum_c S_i^c} F_i \label{eq:orientedcapacity_secondorder}.
\end{align}
\end{subequations}

In the second-order extension of the dynamic system formulation, most of the symbols remain the same, with a few changes:
\begin{defn}[Generic second-order node dynamics system] \label{def:secondorder_ds}
~\\
\begin{itemize} 
  \item Let $\mu \in 2^\mathcal{I}$ (where $\mathcal{I}$ is the set of all input links $i$) be the set of all exhausted input links (this is the same set that was introduced in the first-order solution algorithm in section \ref{sec:firstorder_algorithm}). This is necessary to state the recalculations of supply according to the steps in section \ref{sec:2omimo} when a link exhausts its demand and the net property of a $j^-$ changes.
  \item Paralleling $j^\ast$, let $i^\ast$ denote an exhausted input link.
    An input link is said to be exhausted at time $t$ if $S_{i,j}^c - f_{i,j}^c(t) = 0 \; \forall j,c$.
    Note that the formula for the time of demand exhaustion remains the same as in the first-order case, $T_i = \sfrac{F_i}{p_i}$.
  \item In the first-order dynamic system, we had one continuous state, the node flows $f_{i,j}^c$.
  In the second-order construction, we will need more states to account for the more complicated dynamics.
  We introduce new continuous states $\bar{f}_{i,j}^c(t)$, which are continuous states that are initialized to 0 and have the same continuous dynamics as $f_{i,j}^c$, and $\rho_j^c(t)$, which are the densities of each commodity $c$ in the downstream links $j$.
  To be more specific, we are breaking $f_{i,j}^c(t)$ into two components: $\bar{f}_{i,j}^c(t)$ is the amount of $c$-vehicles that have taken movement $i,j$ since the last mode switch, and $\rho_j^c(t)$ is the (cumulative) density of $c$ vehicles that have entered link $j$ up until the last mode switch.
  We need to do this because the downstream supplies $R_j(t)$ depend on both the commodity mixture of vehicles entering $j$ at time $t$ and the commodity mixture of vehicles that are already in $j$ as of time $t$.
  So, $\bar{f}_{i,j}^c$ will be reset to $0$ after each mode switch.
  This also means that the input flow vehicles that will count against the supplies $R_j(t)$ will be the $\bar{f}_{i,j}^c$.
  \item We assume we have the initial $\rho_j^c(0)$ for all $j,c$.
\end{itemize}

Our hybrid system ($Q, X,$ Init, $\dot{f}_{i,j}^c, \dot{\bar{f}}_{i,j}^c, \dot{\rho}_j^c$, Dom, $\Phi$) is
	\begin{subequations}
  \label{eq:secondorderHS}
	\begin{align}
		Q &= \{  q_{\mu,\nu}\}, \, \mu \in 2^\mathcal{I}, \, \nu \in 2^\mathcal{J} \label{2o:Q} \displaybreak[0]\\
		X &= \RR^{M \cdot N \cdot C} \times \RR^{M\cdot N \cdot C} \times \RR^{M \cdot C} \label{2o:X} \displaybreak[0]\\
		\textnormal{Init} &= Q \times \left\{
		  \begin{aligned}
		  & f_{i,j}^c(t=0)=0 &\forall i,j,c; \\
		  & \bar{f}_{i,j}^c(t=0) = 0 &\forall i,j,c; \\
		  &\rho_j^c(t=0) = \rho_j^c(0) &\forall j,c
		  \end{aligned} \right\} \label{2o:init} \displaybreak[0]\\
		\dot{f}_{i,j}^c &= \begin{cases}
			  \begin{aligned}[c]
			  p_{i,j}
			  \frac{ S_{i,j}^c - f_{i,j}^c(t)
					}{\sum_c \left(S_{i,j}^c - f_{i,j}^c(t)\right)}
			  \bigg(1 - \big| \bigcup_{ \mathclap{
			  \substack{j' \in \nu,\\	\exists \, c: \, f_{i,j'}^c < S_{i,j'}^c }} }
			  \bm{\eta}_{j',j}^i \big| \bigg)
			  \end{aligned}
  			  & \text{ if } i \notin \mu \\[1em]
			  0 &\text{ otherwise}
		  \end{cases} \label{2o:F} \displaybreak[0]\\
		\dot{\bar{f}}_{i,j}^c &= \dot{f}_{i,j}^c \label{2o:xbardot} \displaybreak[0]\\
		\dot{\rho}_j^c &= \frac{\sum_i \dot{f}_{i,j}^c}{L_j} \displaybreak[0]\\
			\textnormal{Dom}(q_{\mu,\nu}) &=
			\left\{ \begin{aligned} 
				f: \, & t \geq T_i \vee \;\forall j,c: \; \bar{f}_{i,j}^c = S_{i,j}^c &\forall i \in \mu, \\
				&t \leq T_i, \;\exists j,c: \; \bar{f}_{i,j}^c \leq S_{i,j}^c &\forall i \notin \mu, \\
				&\sum_i \sum_c f_{i,j}^c = R_j^{q_{\mu,\nu}} &\forall j \in \nu,  \\
					&\sum_i \sum_c f_{i,j}^c \leq R_j^{q_{\mu,\nu}} &\forall j \notin \nu
				\end{aligned} \right\} \label{2o:dom} \displaybreak[0]\\
		\Phi(q_{\mu,\nu},x) &= 
		\begin{cases}
			(q_{\mu,\nu'}, x') &\textnormal{ if } \sum_i \sum_c \bar{f}_{i,j^\ast}^c = R_{j^\ast}^{q_{\mu,\nu}} \\
				&\textnormal{ where } \nu' = \nu \cup j^\ast \\
			(q_{\mu',\nu}, x') &
    			\begin{aligned}[t]
			      \textnormal{ if } &t = T_i \\
			      &\vee \;\forall j,c \; f_{i,j}^c = S_{i,j}^c
		      \end{aligned} \\
				&\textnormal{ where } \mu' = \mu \cup i^\ast.
		\end{cases} \label{2o:r} \\
		\textnormal{where } &x' \triangleq 
		    (\{f_{i,j}^{\prime c}\}, \{\bar{f}_{i,j}^{\prime c}\}, \{\rho_j^{\prime c}\}) = (\{f_{i,j}^c\}, \{0\}, \{\rho_j^c\}) \nonumber \\
		\textnormal{and } &R_j^{q_{\mu,\nu}} \textnormal{from \eqref{eq:mimo_middle}, \eqref{eq:mimoR}, with } w_j = \frac{\sum_c w^c \rho_j^c}{\sum_c \rho_j^c} \nonumber
	\end{align}
	\end{subequations}
	When $\dot{f}^c_{ij} = 0$ for all $i,j,c$, the execution is complete and the node flows are given by the final values of the $f_{i,j}^c$.
\end{defn}

Unsurprisingly, the second-order dynamic system is more complicated than the first-order one.
The reader will note that the discrete dynamics, as discussed before, are triggered by links $j^\ast$ filling and links $i^\ast$ emptying.
The filling of a $j^\ast$ and its entering into $\nu$ remains the same as the first-order system.
The emptying or time-expiry of input links, rather than being encoded in the continuous dynamics as was done in the first-order system's \eqref{general:F}, is now in the discrete dynamics in \eqref{2o:dom} and \eqref{2o:r}.
While it was possible to reduce the number of discrete states in the first-order system by including $i$-emptying in the continuous dynamics in the first-order system, in the second order system, any change in the continuous dynamics changes the output links' $w_{j^-}$, so all continuous dynamics changes must trigger a recomputation of $R_j$, which, in \eqref{eq:secondorderHS}, we do when $\mu$ or $\nu$ change.

Although the second-order system seems much more complex than the first-order system, the second-order solution algorithm is thankfully not that much more complicated than the solution method of the first-order system.
We will see why in the next section.

\subsection{Solution algorithm}
Note that, just as in the first-order system, the second order system has constant continuous dynamics in each discrete state.
This means that, just as in the first-order case, we can easily compute the time that the next discrete state transition occurs.
Like in section \ref{sec:firstorder_algorithm}, this is the smallest of the $t_j$'s and $T_i$'s.
As we said, the input link ``time limits'' remain the same as before, $T_i = (\sum_c S_i^c) / p_i$.
The time that an output link runs out of supply and is filled under the discrete state $q_{\mu, \nu}$, if $t_0$ is the time that the discrete state switched to $q_{\mu, \nu}$ and $j$'s supply was recomputed, is similar to \eqref{eq:modeSwitch4},
\begin{align}
t_j = t_0 + \frac{R_j^{q_{\mu, \nu}}}{\sum_i \sum_c \dot{x}_{i,j}^c} = t_0 + \frac{R_j^{q_{\mu, \nu}}}{\sum_{i \notin \mu} p_{i,j} \bigg( 1 - \big| \bigcup\limits_{ \mathclap{
					\substack{j' \in \nu,\\	\exists \, c: \, f_{i,j'}^c < S_{i,j'}^c }} }
					 \bm{\eta}_{j',j}^i \big| \bigg) } \label{eq:2o_modeswitch}
\end{align}
but differs in two key ways.
First, the term for supply is the recomputed $R_j^{q_{\mu, \nu}}$ from \eqref{eq:mimo_middle} and \eqref{eq:mimoR} (this also accounts for why the numerator in \eqref{eq:2o_modeswitch} does not have a subtracted quantity as in \eqref{eq:modeSwitch4}, as that subtraction of already-filled supply is accounted for in the recomputed supply).
Second is that the denominator is summed over $i \notin \mu$ rather than all $i$, as the set $\mu$ is not in the definition of the first-order dynamic system as stated in section \ref{sec:firstorderds}.

We now state the solution algorithm for the second-order dynamic system.
It follows the same logic as the first-order case: identifying the next $T_i$ or $t_j$ to occur, finding the constant continuous-time dynamics that the system will evolve under until that time, integrating forward in time, a new step of recomputing supply, and repeating.

\begin{defn}[Second-order node model solution algorithm] \label{def:2o_alg}
~\\

\textbf{List of inputs:}
	\begin{itemize}
		\item Per-commodity input demands $S_i^c \; \forall \; i,c$
		\item Per-commodity split ratios $\beta_{i,j}^c \; \forall \; i,j,c$
		\item Per-input priorities $p_i \; \forall \; i$
		\item Per-input link capacities $F_i \; \forall \; i$
		\item Per-movement restriction intervals $\etab_{j',j}^i \; \forall \; i,j,j'$
		\item Per-commodity properties $w^c \; \forall \; c$
		\item Output link fundamental diagram velocity function $v_j = V(\rho_j, w_j) \; \forall \; j$
		\item Initial downstream link per-commodity densities $\rho_j^c(0) \; \forall \; j,c$
	\end{itemize}
\begin{enumerate}
	\item \textbf{Initialize.} \label{step:initialization}
					\begin{alignat*}{2}
			k &= 0 &\begin{minipage}{.4\linewidth}
				\textrm{Set iteration counter to 0}
			\end{minipage} \\
			T_i &= \frac{F_i}{p_i} &\begin{minipage}{.4\linewidth}
				\textrm{Compute input ``time limits''}
			\end{minipage} \\
			S_{i,j}^c &= \beta_{i,j}^c S_i &\begin{minipage}{.4\linewidth}
				\textrm{Compute directed demands}
			\end{minipage} \\
			p_{i,j} &= p_i \frac{\sum_c S_{i,j}^c}{\sum_c S_i^c} &\begin{minipage}{.4\linewidth}
				\textrm{Compute oriented priorities}
			\end{minipage} \\
						\mu(k=0) &= \{i: S_i^c = 0 \; \forall \; c\} &\begin{minipage}{.4\linewidth}
				\textrm{Note initial $(k=0)$ empty input links} 
			\end{minipage} \\[8pt]
			\nu(k=0) &= \{j: V_j( \rho_j, w) = 0\}; \textrm{ where } \rho_j = \sum_c \rho_j^c \quad& \begin{minipage}{.4\linewidth}
				\textrm{Note initial $(k=0)$ zero-supply output links}
			\end{minipage} \\
			f_{i,j}^c(k=0) &= 0 &\begin{minipage}{.4\linewidth}
				\textrm{Initialize all throughflows to 0}
			\end{minipage} \\
			\rho_j^c(k=0) &= \rho_j^c(0) &\begin{minipage}{.4\linewidth}
				\textrm{Initialize all downstream densities to input values}
			\end{minipage}
		\end{alignat*}
												\item \textbf{Compute flow rates.} \label{step:2o_flowrate}
			\begin{align}
		\dot{f}_{i,j}^c(k) = \begin{cases}
			  \begin{aligned}[c]
			  p_{i,j} 
						\frac{ 
				S_{i,j}^c - f_{i,j}^c(t) }{
					\sum_c \left(S_{i,j}^c - f_{i,j}^c(t) \right)}
			  \bigg(1 - \big| \bigcup_{ \mathclap{
			  \substack{j' \in \nu(k),\\	\exists \, c: \, f_{i,j'}^c < S_{i,j'}^c }} }
			  \bm{\eta}_{j',j}^i \big| \bigg)
			  \end{aligned}
  			  & \text{ if } i \notin \mu(k) \\[1em]
			  0 &\text{ otherwise}
		  \end{cases}
		  \quad \forall \; i,j,c \label{eq:flowrate_secondorder}
			\end{align}
		\item \textbf{If} $\dot{f}_{i,j}^c(k) = 0 \; \forall \; i,j,c$, \textbf{then the algorithm ends. Otherwise, continue.}
		\item \textbf{Compute output link supplies.}
		\begin{enumerate}
			\item Compute intermediate states:
		\begin{align*}
			w_{j^-}(k) &= \frac{\sum_c \sum_i w^c(k) \dot{f}_{i,j}^c(k)}{\sum_i \sum_c \dot{f}_{i,j}^c(k)} & \forall \; j \\
			v_{j^-}(k) &= \begin{cases}
			    V(0, w_{j^-}(k)) & \textnormal{if } V(0, w_{j^-}(k)) < V_j(\rho_j(k), w_j(k)) \\
			    V_j(\rho_j(k), w_j(k)) & \textnormal{otherwise}
		    \end{cases} & \forall \; j \\
			\rho_{j^-}(k) &\textnormal{ s.t. } v_{j^-}(k) = V( \rho_{j^-}(k), w_{j^-}(k) ) & \forall \; j
		\end{align*}
		where
		\begin{equation}
			w_j(k) = \frac{\sum_c w^c \rho_j^c(k)}{\sum_c \rho_j^c(k)} \label{eq:wj_k}
		\end{equation}
		\item Compute supplies:
		\begin{align*}
			R_j(k) &=
			\begin{cases}
			    F(w_{j^-}(k)) & \textnormal{if } \rho_{j^-}(k) \leq \rho_c(w_{j^-}(k)) \\
			    \rho_{j^-}(k) v_{j^-}(k) & \textnormal{if }\rho_{j^-}(k) > \rho_c(w_{j^-}(k))
		  \end{cases}
		   & \forall \; j
		\end{align*}
	\end{enumerate}
	\item \textbf{Compute the length of this iteration's timestep.}
	\begin{align}
		dt_j(k) &= \frac{R_j(k)}{\sum_c \sum_i \dot{f}_{i,j}^c(k)}  & \forall \; j \label{eq:dt_j} \\
							dt(k) &= \min\left\{
			 			 \{dt_j(k)\}_{j \notin \nu(k)}, \;
			 \{T_i - t(k)\}_{i \notin \mu(k)} \right\} & \nonumber
	\end{align}
	\item \textbf{Advance forward in time.}
	\begin{align*}
		\Delta f_{i,j}^c(k) &= \dot{f}_{i,j}^c(k) \cdot dt(k) & \forall \; i,j,c \\
		f_{i,j}^c(k+1) &= f_{i,j}^c(k) + \Delta f_{i,j}^c(k) & \forall \; i,j,c \\
		t(k+1) &= t(k) + dt(k) & \\
		\rho_j^c(k+1) &= \rho_j^c(k) + \frac{\sum_i \Delta f_{i,j}^c(k)}{L_j} & \forall j,c
	\end{align*}
	\item \textbf{Update sets of ``completed'' links.} \label{step:completed_links}
	\begin{align*}
		\mu(k+1) &= \mu(k) \cup \left\{i: T_i \leq t(k+1) \right\} \\ 
				\nu(k+1) &= \nu(k) \cup \left\{j: V_j(\rho_j(k+1), w) = 0 \right\}; \text{ where } \rho_j(k+1) = \sum_c \rho_j^c(k+1)
	\end{align*}
	\item \textbf{Set} $k \gets k+1$, \textbf{return to step \ref{step:2o_flowrate} and repeat.}
\end{enumerate}

\end{defn}

\begin{rem}
	Note that in the above, the extra bookkeeping state $\bar{f}_{i,j}^c$ that appeared in our definition of the dynamic system (Definition \ref{def:secondorder_ds} is not necessary.
	This is because, in \eqref{eq:dt_j}, we compute directly the time that the current $R_j(k)$ is exhausted under the flow rates found in Step \ref{step:2o_flowrate}.
	Since the dynamics of $f$ are constant, in practice, when we combine the dynamics \eqref{2o:F} and the switching condition \eqref{2o:dom} (as in \eqref{eq:dt_j}) we can plug the dynamics equation into place and solve for the switching time, eliminating the $\bar{f}$ variable.
\end{rem}

\begin{rem}
	The final downstream link net properties can be computed with \eqref{eq:wj_k} using the final (output) values of $\rho_j^c(k)$.
\end{rem}

% \end{algorithm}
 
\section{Conclusion}
\label{sec:conclusion}
This paper presented a generalization of the widely-used ``Generic Class of Node Model'' macroscopic traffic junction models to the so-called ``General Second Order Model'' flow model.
This paper's results allow the extension of macroscopic modeling of variable-behavior flows (i.e., the macroscopic flow behavior depends on the mixture of behaviorally-different vehicle types) to complex general networks.
Many of these flows and networks had been only able to be modeled by microscopic models that consider the behavioral variability on a per-car level, but macroscopic models that can capture the aggregate features of a more granular model can greatly increase the scale of problems that we are able to study.
As stated before, the second-order flow models have been used to represent flows of great contemporary interest, such as mixtures of human-driven and autonomous vehicles \citep{wang_comparing_2017}.
Researchers and practitioners will need to use every tool available to understand and predict the system-level changes that will arise from the traffic demand changing not just in size, but in characteristics.

Some immediate avenues for future refinement of second-order macroscopic models presented themselves during this paper.
As mentioned in Section \ref{sec:other}, we do not address node supply constraints in this paper's node models.
However, the immediate application of a general, multi-input-multi-output second-order node model, macroscopic simulation of mixed-human-driven-and-autonomous traffic on complex networks, is of particular concern in scheduling problems involving green light timing.
Future work, then, should incorporate the node supply constraints into the general second-order node problem so that they may be used in signal optimization and the still-developing potential that connected and automated vehicles bring to traffic control.

We mentioned briefly in Section \ref{sec:secondorder_intro} that the ARZ model was originally developed as a spatial coarsening of a particular simple car-following model \citep{aw_resurrection_2000,zhang_non-equilibrium_2002}.
More sophisticated car-following models exist.
Some popular ones include the Gipps model \citep{gipps_behavioural_1981}, the so-called General Motors family of car-following models \citep{gazis_car-following_1959}, and the Intelligent Driver Model \citep{treiber_congested_2000} and ``Improved'' Intelligent Driver Model \citep{shen_detailed_2012}.
The Intelligent Driver Model family in particular form the base for many automated cruise control systems in production vehicles today.
That is, the microscopic behavior of many vehicles on the road are well-understood -- either through human-behavioral models or because their behaviors are already (partially) computer-controlled.
There is a need to analytically derive second-order fundamental diagrams in the form of spatial discretizations of contemporary microscopic models, especially in light of recent analyses (e.g., \citep{mehr_how_2019}) that non-controlled deployment of autonomous vehicles can actually worsen traffic network equilibria (i.e., the familiar Braess/Jevons Paradox, but due to changes in the vehicle fleet rather than changes in the road network itself).
In other words, some of the technological development and deployment already exists for macroscopic mixed-autonomy modeling and control, but some of the theoretical understanding is not yet developed (an uncommon situation in engineering, indeed). 
\section*{Acknowledgments}
We thank our colleagues Alex A. Kurzhanskiy and Gabriel Gomes for their reading and feedback.

\bibliographystyle{abbrvnat}

\end{document}